\documentclass[onecolumn]{aa}
\usepackage{graphicx}
\usepackage{txfonts}
\newcommand{\rot}{$v \sin i$ }
\newcommand{\ec}{\'{e}chelle }

\newcommand{\ie}{{\em i.e.,} }
\newcommand{\eg}{{\em e.g.,} }
\newcommand{\cf}{{\em c.f.,} }
\newcommand{\ha}{H$\alpha$ }

\newcommand{\cai}{Ca\,{\footnotesize I} }

\newcommand{\caiihk}{Ca\,{\footnotesize II}~H \&~K }

\newcommand{\lii}{Li\,{\footnotesize I} }

\newcommand{\lsi}{$\mathrel{\hbox{\rlap{\hbox{\lower2pt\hbox{$\sim$}}}\raise2pt
\hbox{$<$}}}$}
\newcommand{\gsi}{$\mathrel{\hbox{\rlap{\hbox{\lower2pt\hbox{$\sim$}}}\raise2pt
\hbox{$>$}}}$}

\begin{document}
   \title{Fundamental properties of pre-main sequence stars in young, southern star forming 
       regions: metallicities}

   \subtitle{The Weak-Lined T-Tauri Stars}

   \author{David J. James
          \inst{1,6}
          \and
          Claudio Melo
          \inst{2,3}
          \and
          Nuno C. Santos
          \inst{4,5}
          \and
          J\'{e}r\^{o}me Bouvier
          \inst{1}}

         \offprints{D. James}

   \institute{Laboratoire d'Astrophysique, Observatoire de Grenoble, 
            BP-53, F-38041 Grenoble, France \\
            \email{david.j.james@vanderbilt.edu,Jerome.Bouvier@obs.ujf-grenoble.fr}
         \and
             European Southern Observatory, Casilla 19001, Santiago 19, Chile \\
             \email{cmelo@eso.org}
         \and
             Departamento de Astronom\'{\i}a, Universidad de Chile, 
             Casilla 36-D, Santiago, Chile  
         \and
             Centro de Astronomia e Astrof\'{\i}sica da Universidade de 
             Lisboa, Observat\'{o}rio Astr\'{o}nomico de Lisboa, Tapada da Ajuda, 
             1349-018 Lisboa,~ Portugal \\
             \email{nuno@oal.ul.pt}
         \and
             Observatoire de Gen\`{e}ve, 51 ch. des Maillettes, CH-1290 Sauverny, 
             Switzerland 
         \and  
             Dept. of Physics \& Astronomy, Box 1807 Station B, Vanderbilt University, 
             Nashville, TN 37235, USA \\
             }

   \date{Received July 24, 2005; accepted ?}

   \abstract{ {\em Aims.}  The primary motivation for this project is to search for 
                           metal-rich star forming regions, in which, stars of 
                           super-solar metallicity will be created, as hopefully, 
                           will be extra-solar planets orbiting them ! The two aims of 
                           this project are: {\bf 1:} to show that our sample stars are 
                           young, lithium rich, magnetically active and non-accreting 
                           kinematic members of their respective regions. {\bf 2:} To 
                           measure the metallicity for such members.

            {\em Methods.} The {\sc feros} \ec spectrograph together with {\sc eso}'s 2.2m 
                           telescope, was used to obtain high resolution (R=32000) spectra 
                           for each of our {\em weak-lined} T-Tauri target stars. The 
                           wavelength range of the spectra is $\simeq 4000-8000$ \r{A}.

            {\em Results.} We find (pre-main sequence) model-dependent isochronal ages of 
                           the Lupus, Chamaeleon and CrA targets to be $9.1 \pm 2.1$ Myr, 
                           $4.5 \pm 1.6$ Myr and $9.0 \pm 3.9$ Myr respectively. The majority 
                           of the stars have Li {\sc i} 6707.8 \r{A} equivalent widths 
                           similar to, or above those of, their similar mass Pleiades 
                           counterparts, confirming their youthfulness. Most stars are 
                           kinematic members, either single or binary, of their regions. We 
                           find a mean radial velocity for objects in the Lupus cloud to be 
                           ${\overline {RV}}=+2.6 \pm 1.8$ km~s$^{-1}$, for the Chamaeleon  
                           {\sc i} \& {\sc ii} clouds, ${\overline {RV}}=+12.8 \pm 3.6$ 
                           km~s$^{-1}$  whereas for the CrA cloud, we find 
                           ${\overline {RV}}=-1.1 \pm 0.5$  km~s$^{-1}$. 

                           All stars are coronally and chromospherically active, 
                           exhibiting X-ray and H$\alpha$ emission levels marginally 
                           less, approximately equal or superior to that of their 
                           older IC 2602/2391 and/or Pleiades counterparts. All bar 
                           three of the targets show little or no signature of accretion 
                           from a circumstellar environment, according to their positions 
                           in a J$-$K/H$-$K$^{'}$ diagram.

                           For the higher quality spectra, we have performed an iron-line 
                           metallicity analysis for five (5) stars in Chamaeleon, four (4) 
                           stars in Lupus and three (3) stars in the CrA star forming regions. 
                           These results show that all three regions are slightly metal-poor, 
                           with marginally sub-solar metallicities, with 
                           $<$[Fe/H]$> = -0.11 \pm 0.14$, $-0.10 \pm 0.04$ \& $-0.04 \pm 0.05$  
                           respectively. 
                           
        {\em Conclusions.} A sample of stars in several nearby, young star-forming regions has 
                           been established, the majority of which is young, lithium rich, 
                           magnetically active and are non-accreting kinematic members of 
                           their respective clouds. Within the errors, each region is 
                           essentially of solar metallicity. The spectroscopic data, comprising 
                           the major complement of observational products for this project, 
                           were collected at the European Southern Observatory at La Silla, 
                           Chile, proposal ID 70.C-0507(A).

   }

\authorrunning{D.J. James et al.}
\titlerunning{Properties of PMS stars in young, southern star forming regions.}
   \maketitle

\section{Introduction}\label{introduction}

Lithium\footnote{The 6708 \r{A} Li feature consists of two line doublets, 
one pair from the $^{6}$Li isotope (6707.933 \& 6708.121 \r{A}), and 
one from the $^{7}$Li (6707.833 \& 6707.932 \r{A}) isotope (Soderblom 
et al. 1990). In the Sun, and meteorites, the isotope ratio 
$^{6}$Li/$^{7}$Li is $\approx8$\% (\cite{anders}), and it is assumed 
that the Li abundance in the Sun, in the field and in young open 
cluster stars ($age$ \lsi 1 Gyr), is dominated by the $^{7}$Li 
isotope.} is a fragile element in the conditions experienced in 
stellar interiors, and is destroyed by $^{7}$Li($p,\alpha)^{4}$He 
and $^{6}$Li($p,\alpha)^{3}$He reactions above stellar temperatures 
$\approx2.5\times10^{6}~$K (\eg Bodenheimer 1965). However, during 
early pre-main sequence [PMS] evolution, $age<5$Myr, solar-type T-Tauri 
stars are fully convective and their central temperatures should not 
yet be sufficiently high to burn Li (Strom 1994). Therefore, the presence 
of appreciable quantities of lithium in the spectra of candidate 
members of young associations and star forming regions [SFRs] is 
a powerful criterion for rejecting field-star non-members. Observations 
of young stars, both with circumstellar accretion disks 
(the so-called {\em classical} T-Tauri stars [CTTS]) and without disks 
({\em weak-lined} T-Tauri stars [WTTS]), have shown that these suppositions 
are generally correct with average lithium abundances of 3.1-3.2 (\eg 
Magazz\`{u}, Rebolo \& Pavlenko 1992; \cite{li-TT1}), which correlates 
well with the {\em cosmic} abundance, \ie the average meteoritic value 
(presumably Li un-depleted - \cite{reeves78}; \cite{anders}; Pinsonneault, 
Kawaler \& DeMarque 1990) and the Li abundance in the interstellar medium 
(\cite{ferlet84}).

Young, rapidly rotating, convective solar-type stars are capable 
of manifesting surface magnetic fields through the interaction of 
rotation, differential rotation and convective motions, ie., the 
dynamo process (Parker 1955, 1979). These induced magnetic fields 
lead to confinement and heating of plasma, the effects of which we 
observe as chromospheric and coronal emissions (\eg Ca\,{\footnotesize II}~H 
\&~K, H$\alpha$ \& X-rays). Empirically, enhanced levels of coronal 
and chromospheric emission are observed in solar-type stars as their 
rotation rate increases (Noyes et al. 1984; Hempelmann et al. 1995), 
although not ad infinitum (James et al. 2000). The correlation is 
founded upon the fact that rapid rotation induces greater dynamo 
action, and hence increased magnetic field production is realized, 
which results in greater magnetically-induced heating. 

During the last decade, the {\sc rosat} satellite has been used 
extensively to perform relatively large X-ray surveys for many 
of the young, nearby SFRs such as Orion and the Taurus-Auriga 
associations (\eg \cite{alc96}; \cite{grosso00}; \cite{stelzer01}), 
with the aim of detecting their members. The reasons for choosing 
the X-ray domain to compliment optical surveys are multi-fold. 
For example, while one cannot fault the effectiveness of utilizing 
the results of large-scale, optical spectroscopic and photometric 
surveys for detecting and characterizing young stellar objects 
in nearby SFRs, such strategies, until recently, required immense 
quantities of telescope allocations and user-intensive 
people-hours. X-ray surveys are far more rapid and efficient. 
This is because the {\em bona fide} SFR members are likely to 
exhibit mean X-ray luminosities far in excess of older-open cluster 
members or field stars of similar mass (\eg \cite{neu95}; 
\cite{stelzer01}), and as such there is a far larger contrast 
in X-rays between true SFR members and background (or foreground) 
field star interlopers. An important caveat to be borne in mind 
ought to be stated. One must remember that while extremely useful, 
X-ray surveys of young SFRs designed to detect substantial fractions 
of their members, will be inevitably be weighted toward finding 
the most rapidly rotating and magnetic active members of these 
regions, and such activity surveys yield inherently biased membership 
samples, which are not fully representative of the evolutionary 
properties of the SFR as a whole. 

Existing studies of metallicity in young SFRs are sparse. Padgett 
(1996) conducted a spectroscopic survey to measure the metallicity 
for a small group of T-Tauri stars in the Orion, Chamaeleon, 
Ophiuchus and Taurus star-forming regions. She concluded that 
the mean iron abundance derived for these four SFRs is roughly 
solar, albeit with a dispersion (error on the mean metallicity) 
of about 0.05-0.06 dex. Despite being very interesting, Padgett's 
study only included a few stars per cloud (typically 5-8 stars) 
which slightly weakens her conclusions. She herself comments 
that her results must be confirmed on the basis of a larger 
sample and higher $S/N$ spectra. Some mention of a spectroscopic 
survey yielding metallicities for young T-Tauri stars is also 
made by Keller \& Koerner (2003), in support of a {\sc sirtf} 
Legacy program. However to date, no such results have been 
forthcoming in the literature. 

Our current research project takes advantage of extant X-ray and 
optical surveys of young, southern SFRs so that we may investigate 
the stellar properties of a sample of WTTSs found within them. 
Our initial mission goals are simple. [{\bf 1}] To define and 
catalogue a subset of genuine, {\em bona fide} members of the parent 
associations. [{\bf 2}] To refine the sample further in order to 
flag multiple systems for separate analyses. [{\bf 3}] To ensure 
that the remaining single, genuine members are slowly-rotating 
(\lsi 30 km~s$^{-1}$ say). This 'clean' sample facilitates the 
following scientific aims. 

First, to determine the metallicity of young solar-type WTTSs 
in different nearby SFRs. This is because recent studies have 
shown that stars hosting Jupiter-like planets tend to be more 
metal rich (by say 0.25 dex in [Fe/H]) than those star systems 
for which no extra-solar planets have been found (Gonzalez et 
al. 2001; Santos, Israelian \& Mayor 2001, Santos et al. 2004). 
Pursuant to these metallicity measurements, our future research 
campaigns will be geared toward searching for, and then 
characterizing, extra-solar planets orbiting these young stars, 
focusing our search efforts on more metal-rich SFRs.

Second, a binary population can be identified for follow-up 
photometric campaigns, with the goal of discovering eclipsing 
binaries in the SFRs. Such systems would permit us to determine 
the intrinsic distances to each SFR and measure empirical masses 
directly in young, pre-main sequence star-forming environments. 
This will allow us to probe cluster characteristics, such as 
luminosity and mass functions, in a model independent way.

Third, if the samples are sufficiently large, we shall ascertain 
if there exists sufficient empirical evidence which indicates 
that stellar metallicity can affect the global properties of 
young stars, such as rotation, lithium abundance, magnetic 
activity manifestations and multiplicity. Such a study is 
invaluable if one is finally able to judge whether environment 
and/or initial conditions have r\^{o}les to play in controlling 
the evolution of stellar parameters such as angular momentum 
and surface lithium abundances as stars evolve onto the main 
sequence.

This first article of a series, outlines our initial refinement 
of a sample of WTTSs in several southern SFRs for which we have 
obtained high resolution, high-S/N, optical spectroscopic 
observations. A description of the observations and their 
reduction are detailed in \S~\ref{target-obs}. A presentation 
of the spectroscopic results is presented in \S~\ref{results} as 
well as the analyses we have used to eliminate SFR non-members, 
multiple stars and those systems which we do not, in hindsight, 
adjudge to be true WTTSs. A presentation and discussion of the 
metallicity analysis we have performed for {\em bona fide}, young, 
single, non-accreting members of each SFR is given in \S~\ref{metal}.

\section{Target Selection and Observations}\label{target-obs}

WTTS catalogues for each SFRs we are studying were constructed from 
{\sc rosat} All-Sky Survey [{\sc rass}] detections in and around 
these SFRs. The rotation-magnetic activity paradigm will of course 
play a r\^{o}le, and we thus expect our sample to be mostly probing 
the tail of the angular momentum distribution where the rapid 
rotators are situated, as well as the binary/multiple systems. 

For each SFR, the WTTS sample was constructed from {\sc rass} 
detections in and around the SFR to satisfy at least one of 
the following criteria: ({\bf I}) those stars having spectral 
types of G and early$-$K. ({\bf II}) those stars having visual 
magnitudes of 12 or brighter ; ({\bf III}) those stars with weak 
signatures of infra-red excesses (and magnetic activity), \ie 
having H$\alpha$ EWs $<<$ 10 \r{A} ; ({\bf IV}) those stars 
exhibiting substantial Li {\footnotesize{I}} 6707.8 \r{A} EWs, 
which is indicative of youth. ({\bf V}) those stars which are 
not components of multiple systems, as judged from existing 
kinematic data.  

These criteria were chosen so that each sample has the highest 
probability of representing a {\em bona fide} set of solar-type 
members of the SFR, and ensuring that the stars are indeed of 
the weak-lined T-Tauri class without active accretion signatures. 
Of course, one may also utilize infra-red colours to assist in 
the elimination of field-star interlopers and CTTSs from the 
WTTS sample. This is because the WTTSs should show negligible 
evidence of infra-red colour excesses. Such excesses may be 
attributable to CTTS-like systems having retained their 
circumstellar accretion disks (see \S~\ref{results} for further 
details). In the process of the current research, this analysis 
was not carried out until after our spectroscopic observing run, 
as we were unsure of the data quality status of the two Micron 
All-Sky Survey [{\sc 2mass}].\footnote{This publication makes use of data 
products from the Two Micron All Sky Survey, which is a joint 
project of the University of Massachusetts and the Infrared 
Processing and Analysis Center/California Institute of 
Technology, funded by the National Aeronautics and Space 
Administration and the National Science Foundation.} Under 
such a cloud of uncertainty, we thus preferred to attend 
the release of the {\sc 2mass} all-sky data release in March 2003. 

\begin{figure}
   \centering
   \includegraphics[angle=-90,width=\textwidth]{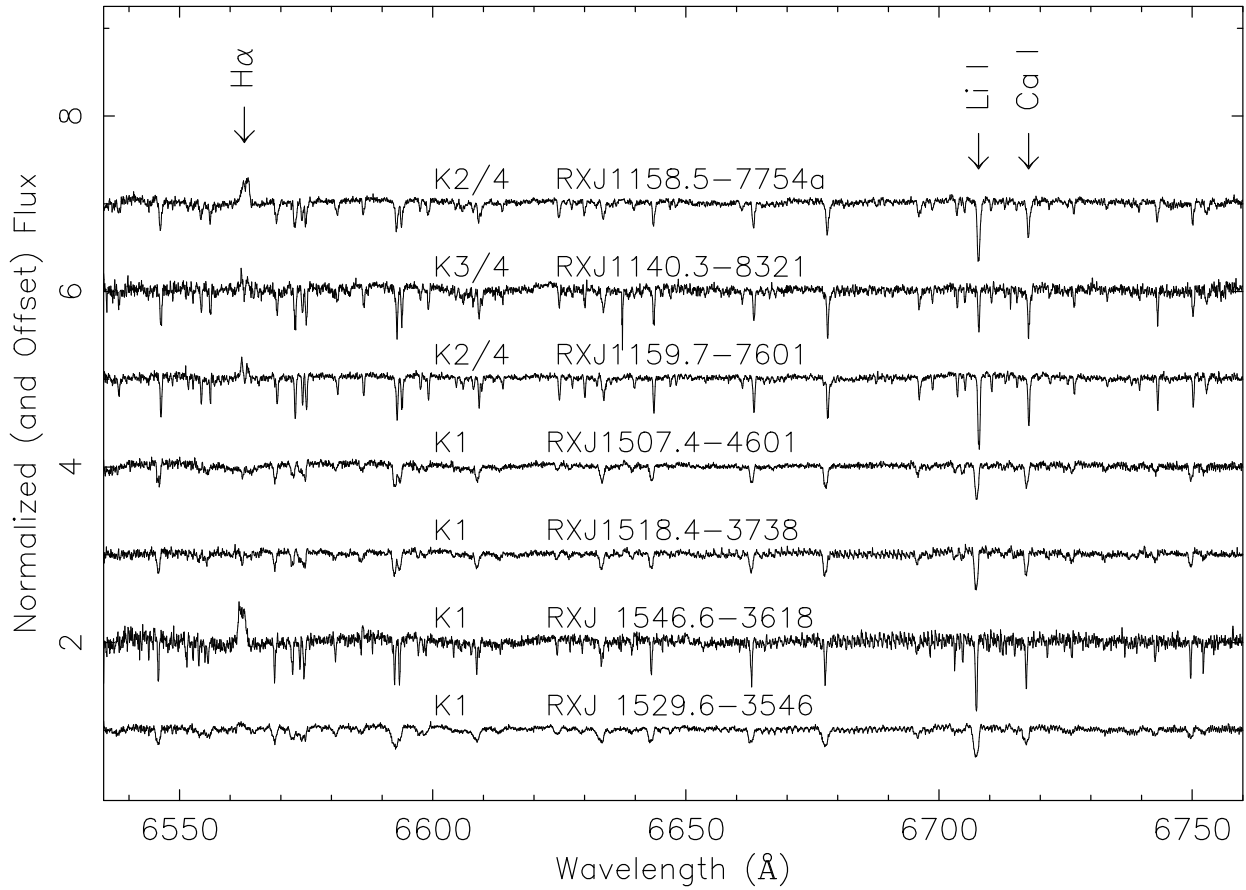}
      \caption{Example normalized {\sc feros} spectra are presented 
            for a subset of the WTTS candidates in the SFRs 
            under study. The spectra are offset for clarity. 
            Several lines of astrophysical interest are 
            annotated, such as the Balmer H$\alpha$ line at 
            6562.8\r{A} as well as the \lii (6707.8\r{A}) and 
            \cai (6717.7\r{A}) lines. }
         \label{FEROS}
   \end{figure}

Each of our WTTS candidates for each SFR was observed at high 
resolution using the {\sc feros} \ec spectrograph at the coud\'{e} 
station of the 2.2m telescope (fork-mounted, Ritchey-Chr\'{e}tien) 
situated at the European Southern Observatory [ESO], La Silla, 
Chile during the nights of 13, 14, 19 \& 20 March, 2003. The 
observations were performed using a 79 lines mm$^{-1}$ \ec grating 
and an {\sc eev} 2K$\times4$K CCD as detector, with 2.7 arcsec sky 
and target fibres. This set-up yielded a {\sc fwhm} of cross-correlated 
ThArNe arc lines of 0.17 \r{A} at 5500 \r{A}, and a useful wavelength 
range of $\simeq 4000-8000$ \r{A}. Examples of the processed spectra 
in the vicinity of the H$\alpha$ region of the spectrum are shown 
in Figure~\ref{FEROS}.

The {\sc feros} spectra were used to obtain heliocentric radial 
velocities [RVs] and projected equatorial rotational velocities 
[$v \sin i$] by using cross-correlation techniques (Tonry \& Davis 
1979) in concert with high S/N, IAU RV standards and slowly rotating 
stars of similar spectral type to the targets, using the spectral 
range $5420-5620$ \r{A}. This spectral order yields spectra containing 
many metal absorption lines and little telluric contamination. Radial 
velocity zero points were set by reference to spectra of the IAU 
radial velocity standard stars HR 1829, HR 2701, HR 4540, HR 5384, 
HR 6349 \& HR 6468. Cross-correlation of each standard star spectrum 
with those of the other RV standards revealed that external errors 
on the standard system were about $0.2-0.4$ km~s$^{-1}$. Random 
errors due to the poorer S/N of the target spectra, spectral-type 
mismatch between target and standard and the effects of broadening, 
were determined by multiple, Monte Carlo-like simulations. Errors 
varied from 0.2 km~s$^{-1}$ for slowly rotating stars, increasing 
to $\simeq 2-3$ km~s$^{-1}$ for targets rotating at $\geq 20$ 
km~s$^{-1}$.

The {\sc fwhm} of the cross-correlation peaks obtained with slowly 
rotating, inactive standard star spectra of similar spectral types 
to the targets, were measured to provide \rot determinations. The 
standard stars were chosen to have minimal activity in the 
chromospheric \caiihk lines, with rotation periods estimated from 
a correlation between rotation and chromospheric activity 
(Rutten 1987). The relationship between {\sc fwhm} of the 
cross-correlation peak and \rot was calibrated by convolving 
the standard star spectra with limb-darkened ($\varepsilon=0.6$) 
rotational broadening profiles (\cite{gray}) and cross-correlating 
with the unbroadened templates. Random errors were calculated 
by multiple, Monte-Carlo-like simulations, and proved extremely 
sensitive to S/N and rotation. 

The error simulations took the form of testing the random effects 
of S/N, rotation and spectral type mis-match between target and 
standard star spectra. In the first instance, we took a very high 
S/N, narrow-lined standard star spectrum and randomly degraded it 
to a much lower S/N (we choose S/N=5,10,20,40,50,100) many times 
(normally 100 lower quality spectra were created). By 
cross-correlating these degraded spectra with the original 
high-S/N spectrum and noting the variation in cross-correlation 
peak {\sc fwhm} values, the random error due to spectral quality 
can be inferred. Like-wise for the effects of target rotation 
upon the random error budget, we followed a similar procedure 
as before, except in this instance, each of the degraded spectra 
were rotationally broadened (artificially with a Gray profile) 
to various rotation rates. Again, by cross-correlating suites of 
rotationally broadening, randomly degraded spectra against the 
original {\em clean} spectrum, line-broadening and spectral 
quality errors can be inferred. Finally, to estimate errors 
due to spectral mis-match, high quality, narrow-lined spectra 
of standard stars, of various spectral types, were cross-correlated 
against each other to catalogue variations in cross-correlation 
peak {\sc fwhm} values.

Given that the majority of the {\sc feros} data obtained for 
this programme yield spectra with a S/N \gsi 50, and we obtained 
spectra of standard stars with spectral-types matching those 
of our targets, it is likely that our \rot measurements are 
accurate to about $10\%$, down to a lower limit of 6 
km~s$^{-1}$ (the instrumental resolution of {\sc feros} is 
probably considerably lower, possibly as low as 2 km~s$^{-1}$, 
e.g. see Santos et al. 2002; Melo, Pasquini \& Medeiros 
2001, however we choose a more conservative velocity limit 
of 6 km~s$^{-1}$). Such conservatism does not alter the 
scientific conclusions resulting from this study.

For each target spectrum, we have also measured the EWs of the 
H$\alpha$ line at 6563\r{A} and the Li {\sc i} 6708\r{A} resonance 
lines using both the direct integration and the Gaussian 
fitting methods. For the lithium region, we simply rectified 
the spectrum before measuring the \lii EW, and so our values 
include the contributions from the small Fe {\sc i}+CN line at 
6707.44 \r{A}, leading to measured \lii EWs that are representative 
of a slightly ($10-20$ m\r{A}) overestimated photospheric Li presence. 
For instance, Soderblom et al. (1993a) report that this Fe line blend 
has an EW = [20(B-V)$_{0}$ - 3]m\r{A}, for main sequence, solar-type 
stars. In the case of H$\alpha$, the normalized H$\alpha$ spectrum 
of a minimum-activity standard star (old, Li depleted, 
slowly rotating stars - {\em viz} HD 36436 [F7V], HR 5384 [G1V], 
HD 65216 [G5V], HD 73256 [G8V], HD 22049 [K2V], HD 160346 [K3V] \& 
HD 156026 [K5V]), closest in spectral type to that of each 
target, was shifted to each target's RV, artificially 
broadened to the target star's \rot (except in those cases where 
\rot$<10$ km~s$^{-1}$) and subtracted from the target's 
normalized H$\alpha$ spectrum. This procedure yields the 
residual \ha emission of a target star. In effect, we are 
removing a photospheric contribution from the stellar \ha 
profiles which we are measuring. These \ha residuals represent 
the {\em filling in} of the core of the \ha profile, relative to 
the similar-mass, standard star spectrum, which we assume to 
be the result of dynamo$-$induced chromospheric activity and/or 
an accretion signature if the star is of the {\em classical} type.

At this point several cautionary notes should be sounded. There 
may be systematic errors introduced in the measurement of \ha EWs 
using the {\em spectral subtraction} technique described above, however 
they should not be so large as to prevent us from classifying the 
targets as WTTSs or CTTSs. First, if the metallicity of the target 
is significantly different from that of the standard star, then 
there may be an offset between the two spectra due to a difference 
in the local continuum flux and in the temperature and conditions 
under which a specific line is formed. Second, if any part of 
the line contribution was formed in an optically thick region of 
a target star, the entire residual EW of a spectral line cannot 
have its origins in regions of optically thin chromospheric emission. 
This is because Compton scattering from more energetic photons 
(than the line) may contribute to the line flux, which is herein 
assumed to be chromospheric in origin. Third, it may be possible 
that these PMS stars, some with quite high rotation velocities, 
may have activity cycles similar to those observed on the Sun, 
and may be more chromospherically active during the current 
observing season than, say, some time in the past (on timescales 
of rotation periods to years). Fourth, we choose not to telluric 
correct the \ha profiles of the targets. The subtraction of a 
similar spectral-type standard star \ha profile will in essence 
help to mitigate this by subtracting out the telluric feature, 
albeit at a strength when the standard star spectrum was taken. 
Some differential residual telluric feature may result after 
spectral subtraction, however we expect that this contribution 
will be no more than $10-20$m\r{A}. This is likely to be of 
the same order as the internal statistical error bars on the 
EW measurement (assumes $10\%$ errors, as is common for 
intermediate S/N, high resolution spectra).

\section{Results}\label{results}

The complete astrometric and photometric data ensemble for the input 
catalogue for each SFR is detailed in Table~1. Segregated 
into specific SFRs, the data presented represent the target {\sc rass} name 
(col 1), its alternative name if available (col 2), its optical 
position taken from Digitized Sky Survey [{\sc dss}] images (cols 3 \& 4 - 
precise to $\simeq 1$ arcsecond - RMS), the optical magnitudes, colours, 
mass indices in the form of  V, V$-$I \& spectral type (taken from the 
literature; cols $5-7$) and their {\sc 2mass} JHK$^{'}$ magnitudes and errors 
of any detections within 2 arcseconds of the optical positions (cols $8-10$).

The cross-identifications and optical data sources for individual 
members of those regions detailed in Table~1 are thus: 
for the Lupus SFR they are Krautter et al. (1997 - {\sc k} sources) \& 
Wichmann et al. (1997a, b), whereas for Chamaeleon regions {\sc i} \& 
{\sc ii} they are Alcal{\'a} et al. (1995), Covino et al. (1997) and 
Alcal{\'a} et al. (2000). We also used ID/optical data for the CrA SFR 
from Neuh\"{a}user et al. (2000), and the few data we use for the Rho 
Oph cloud are from Mart\'{\i}n et al. (1998).

The results of the spectroscopic kinematic analyses and a simplistic 
treatment of the X-ray properties of our sample stars are presented in 
Table~2. In the first instance, a radial velocity analysis 
permits one to establish firm membership constraints. Given that all 
the candidates are likely to be cluster members, because of the 
initial selection criteria for the sample (see \S~\ref{target-obs}), 
the primary use of the RV data is thus to establish SFR membership 
and multiplicity for each star. In the worse case scenario, we can 
also establish cluster {\em non-membership}.

Again, to allow husbandry of all the X-ray/kinematic data for each 
SFR, we categorize the SFR samples in the same way as was performed 
in Table~1. For each target, its {\sc rass} X-ray count rate 
and luminosity (for the {\sc rosat} passband of $0.1-2.4$ keV) 
is catalogued in columns $2-3$ of Table~2. The standard 
{\sc rass} conversion factor, in translating X-ray count rate to X-ray flux 
for negligible interstellar absorption, of $6\times10^{-12}$ erg 
cm$^{-2}$ count$^{-1}$ has been used to determine the X-ray flux of 
each target. Of course, the tabulated fluxes are most likely lower 
limits, as one would find it hard to believe that some level of 
absorption of X-ray flux was not taking place in the local SFR 
environment. X-ray luminosities, as ratios to bolometric 
luminosities, are provided in column 4. SFR distances are 
detailed in the table footnotes and bolometric corrections 
have been calculated for each star, as a function of 
spectral type, using the relationships reported by Kenyon 
\& Hartmann (1995 - [KH95]). Interstellar reddening estimates 
have been used from the data given in Table~4 (see 
also \S~\ref{HRD-anal}). In columns 5 \& 6, we report the literature 
values (see Table~2 footnotes for references) of the 
angular momentum parameters photometric rotation period \& \rot 
for each target, if known. Finally, our kinematic results for 
each star observed during this campaign, namely the heliocentric 
Julian data [HJD], radial velocity and $v \sin i$,  are presented 
in columns $7-9$.  For completeness, we also add a comments column 
in which we detail target RVs found in the literature and/or whether 
a target has been found to be a double-lined spectroscopic binary 
[SB2] system. 

Two rapidly rotating Chamaeleon stars, RX J0850.1-7554 \& RX 
J0951.9-7901, yielded cross correlation functions, when 
their spectra were cross correlated against the suite of 
RV standard star spectra, which appeared somewhat double-peaked 
$-$ indicative of an SB2 system. These cross correlation peaks 
were not always well separated, nor extremely clear, and we 
concede that these stars may be single in nature. Their rapid 
rotation and/or the presence of surface star-spots may be the 
cause of such inhomogeneities or double-peaks in the cross 
correlation functions, and not because of a secondary stellar 
component. We label them as {\em SB2?} systems (possible SB2s) 
and urge caution in their interpretation. One star in 
Chamaeleon, RX J1303.5-7701, which is an RV non-member, was 
remarkable in that its spectrum looked like that of a hot, 
high mass star of early type, and is in all likelihood a 
background reddened giant.

The spectral analysis data relating to the Li {\sc i} 6708\r{A} \& 
Balmer H$\alpha$ 6563\r{A} lines for each SFR target are catalogued 
in Table~3. The target identifier and the HJD of 
observation are listed in columns 1 \& 2, as well as the approximate 
signal-to-noise [S/N] ratio of the target's spectrum, at the nearest 
continuum region to the 6708\r{A} line, in column 3.  As a sanity 
check, we choose to measure target Li {\sc i} 6708\r{A} EWs using the 
direct integration method and the Gaussian fitting method. These 
data are presented in columns 4 \& 5. In all cases, these two measures 
are within $10\%$ of each other. In column 6, we list our adopted 
Li EWs, which are based on the Gaussian fitted line EWs, and a 
$10\%$ error budget. We have chosen these EWs preferentially because 
if an individual spectrum suffers from a nearby cosmic ray event, 
a Gaussian fit to the line profile is less likely to be affected 
(unless the cosmic ray destroys several pixels, $>5$ say) than 
by integration. If a target has two sets of Li {\sc i} EW from 
different epochs of observation, its adopted EW is the weighted 
mean of the two Gaussian-fitted EWs, weighted by the S/N of each 
of its spectra. If there already exists an extant Li measure for 
a target in the literature, then it is detailed in column 7 
(see RV references in Table~2 for citations). Finally, 
the Gaussian-fitted and direct integration H$\alpha$ EWs are 
presented in columns 8 \& 9 for the {\em residual} chromospheric-emission 
profiles we have created. Extreme chromospheric emission, rapid 
rotation and/or surface star-spots can deform the shape of spectral 
line profiles, especially ones that have contributions from magnetic 
heating processes. Therefore, in this instance, we choose to use 
the direct integration EW for the H$\alpha$ line, and not the 
Gaussian fitting method one, to mitigate against cases where the 
residual emission profile is distorted. We note in passing that 
the observation of the Lupus star RX J1605.7-3905 yielded a visually 
bizarre spectrum around the Li {\sc i} region, and so we present 
only its Gaussian fitted EWs.

To attain our science goals, the primary task undertaken was 
to establish membership of the SFR for each WTTS candidate. 
Combined with the initial selection criteria of the samples, 
outlined in \S~\ref{target-obs}, we also use several physical 
characteristic criteria of the targets to ensure that our 
sample really is representative of a ensemble of young, single 
WTTSs, and does not include binary members and/or field star 
interlopers nor stars of the classical T-Tauri class. 

And so, the results section is structured thus: ({\bf a}) a 
discussion of the radial velocity data is presented to establish 
kinematic {\em membership} of each SFR, either single, binary or 
non-member in nature; ({\bf b}) establishing fundamental parameters 
for each target, such as luminosity and effective temperature, 
for comparison to theoretical PMS isochrones and mass tracks. An 
age analysis to confirm youth is performed; ({\bf c}) a comparison 
of how much lithium each PMS star has in its atmosphere is made 
with the members of the Pleiades cluster, an open cluster of 
known-age ($\simeq 100$ Myr). Such a comparison allows us to further 
verify that the {\em members} of each SFR are indeed young; ({\bf d}) 
an analysis is performed for each star to show that it does {\bf not} 
display considerable H$\alpha$ emission nor has considerable 
infra-red excess, so that we can confirm that each SFR {\em member} 
is of the {\em weak-line} class. If a sample of stars can be created 
for each SFR which is able to satisfy the membership, youth and 
WTTS-class criteria, a metallicity study can subsequently be 
performed upon their spectra.

\subsection{Radial Velocities:} 

For each of the WTTS candidates listed in Table~1, 
we present at least one radial velocity measure of our own 
(and its HJD), and indeed, in many cases two measurements. Used 
in concert with any extant RV measurements to be found in the 
literature, we feel confident that we are able to ascertain 
whether candidates are probable kinematic members (single or 
binaries) of their respective SFRs. Of course, using this 
criterion, we cannot yet say whether the candidate is young 
or is a {\em weak-lined} or {\em classical} T-Tauri star.

The comparative use of a kinematics criterion is fortunately 
available from existing spectroscopic surveys of SFRs. For 
instance, Wichmann et al. (1999) show that the mean radial 
velocity for an ensemble of 49 X-ray selected, Li-rich, 
late-type WTTS stars in the Lupus SFR is 
${\overline {RV}}=+2.72 \pm 1.15$ km~s$^{-1}$. For the 
Chamaeleon SFRs, Colvino et al. (1997) have shown that 
their Li-rich, {\sc rass} detected, WTTS stars have RVs in 
the range $+12 <$RV$< +18$ km~s$^{-1}$, with a clearly 
defined peak in the RV histogram at $\simeq +13$ km~s$^{-1}$ 
(their figure 7). Moreover, the results of Walter et al. 
(1997) and Neuh\"{a}user et al. (2000) show that the majority 
of their Li-rich T-Tauri candidate members in the Corona 
Australis (CrA) SFR have RVs in the range $0\rightarrow -2$ 
km~s$^{-1}$. In addition, a recent multiplicity study, by 
Melo (2003), of young, T-Tauri stars in several SFRs further 
corroborates these systemic kinematic results, as have been 
reported by the authors cited above. In any case, the bulk 
1-d kinematic properties of the SFRs that we are studying 
are now reasonably well-defined.

\begin{figure}
   \centering
   \includegraphics[angle=-90,width=\textwidth]{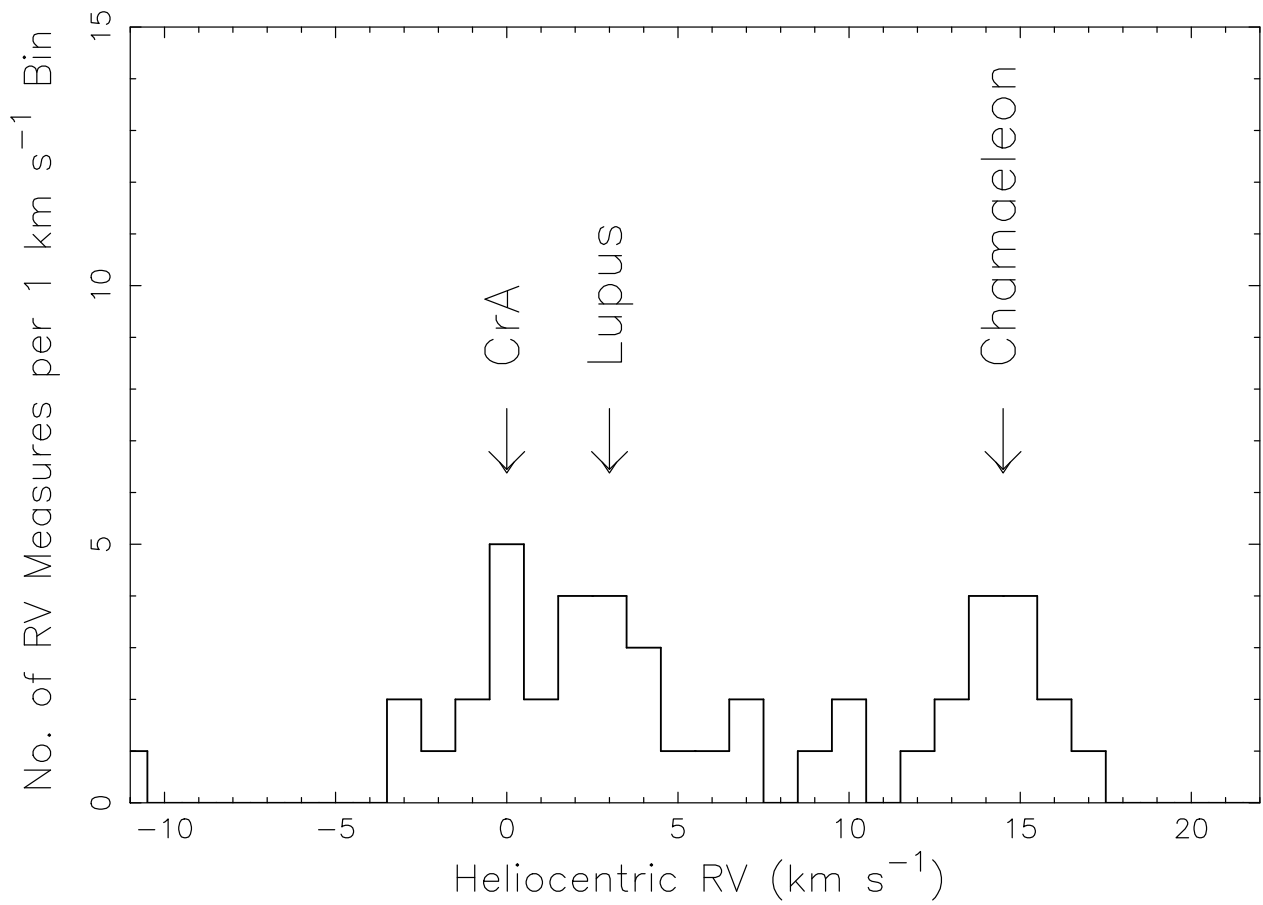}
      \caption{An RV histogram is presented for a subset of all 
           of the RV measurements determined during the course 
           of this study. Peaks in the RV histogram, which are 
           co-incident with the published cluster centric RVs 
           of various SFRs, are labelled. }
         \label{RV-hist}
   \end{figure}

In constructing a radial velocity histogram one can search 
for local peaks in the kinematic space density. Starting 
from the RV measurements detailed in Table~2, 
such an analysis is presented in Figure~\ref{RV-hist}. 
While our data ensemble, on a region by region basis, 
is somewhat limited by small number statistics, it can 
surely not be aleatory that the three peaks in the RV 
distribution are coincident with the mean 1-d kinematics 
of the CrA, Lupus and Chamaeleon SFRs discussed above.

In fact, if one were to take the average of each star's RV 
measure, for each SFR, as is detailed in column 9 of 
Table~2, one would obtain a mean RV for objects 
in the Lupus cloud to be ${\overline {RV}}=+2.6 \pm 1.8$ 
km~s$^{-1}$, for the Chamaeleon {\sc i} \& {\sc ii} clouds, 
${\overline {RV}}=+12.8 \pm 3.6$ km~s$^{-1}$ whereas for 
the CrA SFR, ${\overline {RV}}=-1.1 \pm 0.5$ km~s$^{-1}$. 
Such an analysis includes all stars in each SFR except 
for RX J1307.3-7708 (which is a probable non-member of 
the Chamaeleon region) and the Chamaeleon object 
RX J1303.5-7701. The former is an RV \& photometric 
non-member of the Chamaeleon SFR, and is a low lithium 
object. The latter is an RV non-member and appears in 
our spectra to be of early spectral-type, and we 
henceforth classify RX J1303.5-7701 as a field 
interloper in the star field of Chamaeleon, and we 
shall not consider this star further. We issue a 
{\em caveat lector} for this star because it has 
previously been classified as a G7 star 
(Alcal{\'a} et al. 2000), in discord with our 
observations.

The majority of our observed target stars are most likely 
{\em bona fide} kinematic members of their respective SFRs. 
Of course, while we were only able to record one or two epochs 
of velocities for the SB2s, the RVs of their constituent 
components appear to straddle the mean RV of their respective 
associations, indicating to us that they have a reasonable 
probability of being members. For the two Rho Oph stars, the 
only real inference that we can make is that their two 
single-epoch RV measures are consistent with each other. In 
any case, these two Rho Ophiuchus stars have infra-red colours 
consistent their being {\em classical} T-Tauri stars, and 
we do not consider them further (see below). 

\subsection{Stellar Luminosity, Lithium Depletion and Stellar Youth:} \label{youth}

To discriminate against field star interlopers having radial 
velocities matching those of the parent SFR under study, thus 
mimicking true SFR members, one useful characteristic we can 
exploit is stellar age (\cf \S~\ref{introduction}). In 
judging stellar youth for our sample, we consider two 
analyses; ({\bf {\sc i}}) comparison of targets' stellar 
luminosity-effective temperature data and theoretical 
stellar models on an Hertzsprung-Russell diagram [HRD]. 
({\bf {\sc ii}}) comparison of targets' lithium content 
with that of a cluster of known age. 

\subsubsection{An Hertzsprung-Russell Diagram:} \label{HRD-anal}

Determination of fundamental stellar parameters for each SFR 
target has been achieved by converting observational parameters 
such as magnitudes and spectral types to luminosity and effective 
temperature using statistically-large observational samples 
of Galactic field stars and theoretical stellar models. Such 
data for each SFR candidate, where known, are tabulated in 
Table~4. Effective temperatures and bolometric corrections 
have been calculated for each star, as a function of spectral 
type, using the relationships reported by KH95. Interstellar 
reddening is calculated by comparing target V$-$I colours 
(or in five cases, their J$-$H colours) to KH95 theoretical 
colours derived from spectral type. The SFR distances adopted 
for the luminosity calculations are the same as those cited 
in Table~2.

\begin{figure}
   \centering
   \includegraphics[angle=-90,width=\textwidth]{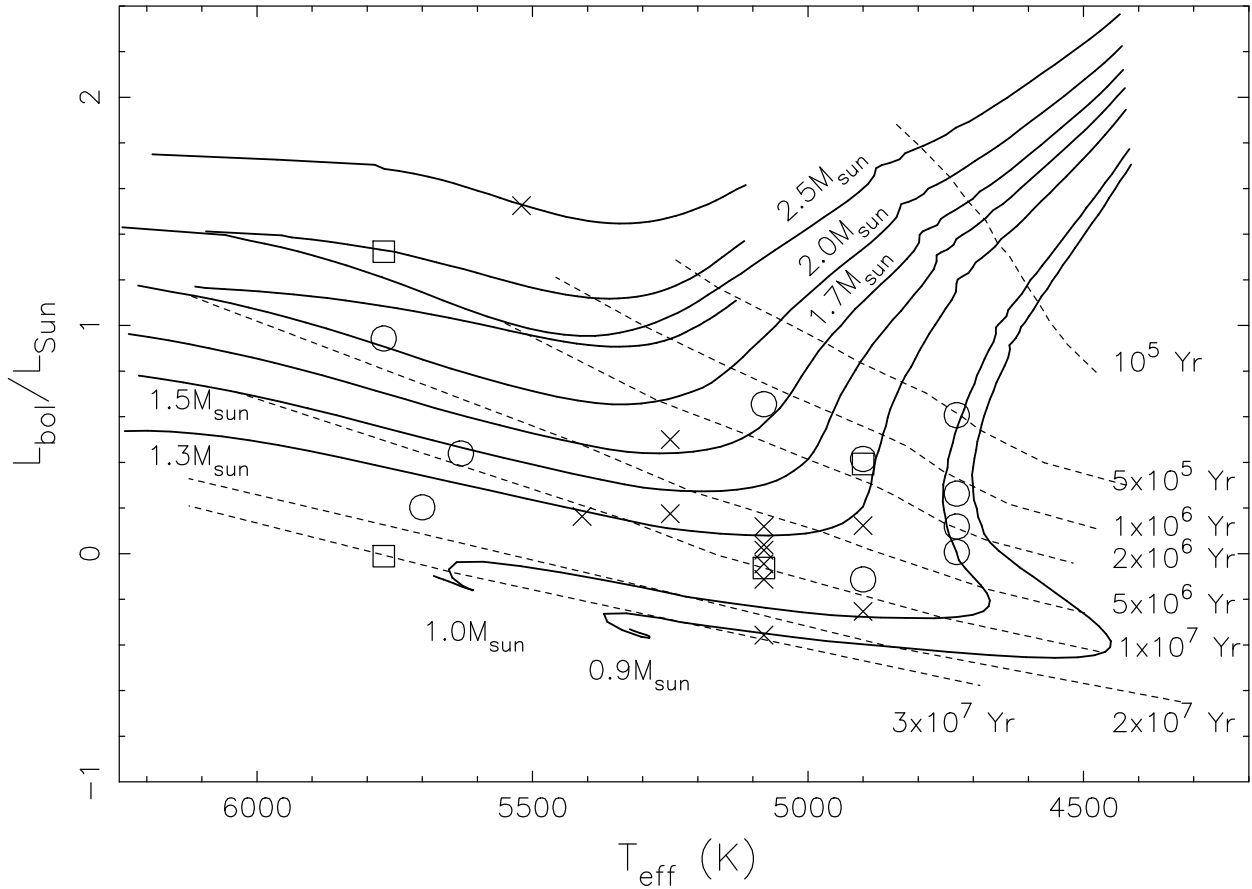}
      \caption{An Hertzsprung-Russell diagram, which includes 
           data points representing our target SFR candidate 
           members, is plotted. Stellar isochrones 
           (dashed lines) and mass tracks (solid lines) are 
           also plotted, and are taken from the theoretical 
           pre-main sequence, solar-metallicity stellar models 
           of D'Antona \& Mazzitelli (1997). The Lupus SFR 
           data are depicted by the central crosses, open 
           circles represent the Chamaeleon data, whereas 
           the open squares are for the CrA SFR data.}
         \label{data-HRD}
   \end{figure}

Bolometric luminosities, as a fraction of solar luminosity, 
and effective temperatures for each SFR target have subsequently 
been plotted on an HRD, which is shown in Figure~\ref{data-HRD}. 
Solar metallicity (with an initial deuterium abundance of 
$4.5\times10^{-5}$), pre-main sequence isochrones and mass 
tracks, computed by D'Antona and Mazzitelli (1997), are also 
plotted. By comparison to these theoretical models, we have 
determined a stellar mass and age for each target star, for 
which we were able to determine a luminosity. Such fundamental 
data are reported in last two columns of Table~4. 

Reassuringly, with the exception of a small handful of stars, 
the majority of the SFR candidate members have isochronal 
ages considerably younger than 20 Myr. In fact, for the Lupus, 
Chamaeleon {\sc i} \& {\sc ii} and CrA SFRs, the mean isochronal 
ages are $9.1 \pm 2.1$ Myr, $4.5 \pm 1.6$ Myr and $9.0 \pm 3.9$ 
Myr respectively. These values agree remarkably well with those 
summarized by Rebull, Wolff \& Strom (2004 - their table~2) and 
Neuh\"{a}user et al. (2000).


\subsubsection{Detection of Lithium:} \label{lithium}

\begin{figure}

   \centering
   \includegraphics[angle=-90,width=\textwidth]{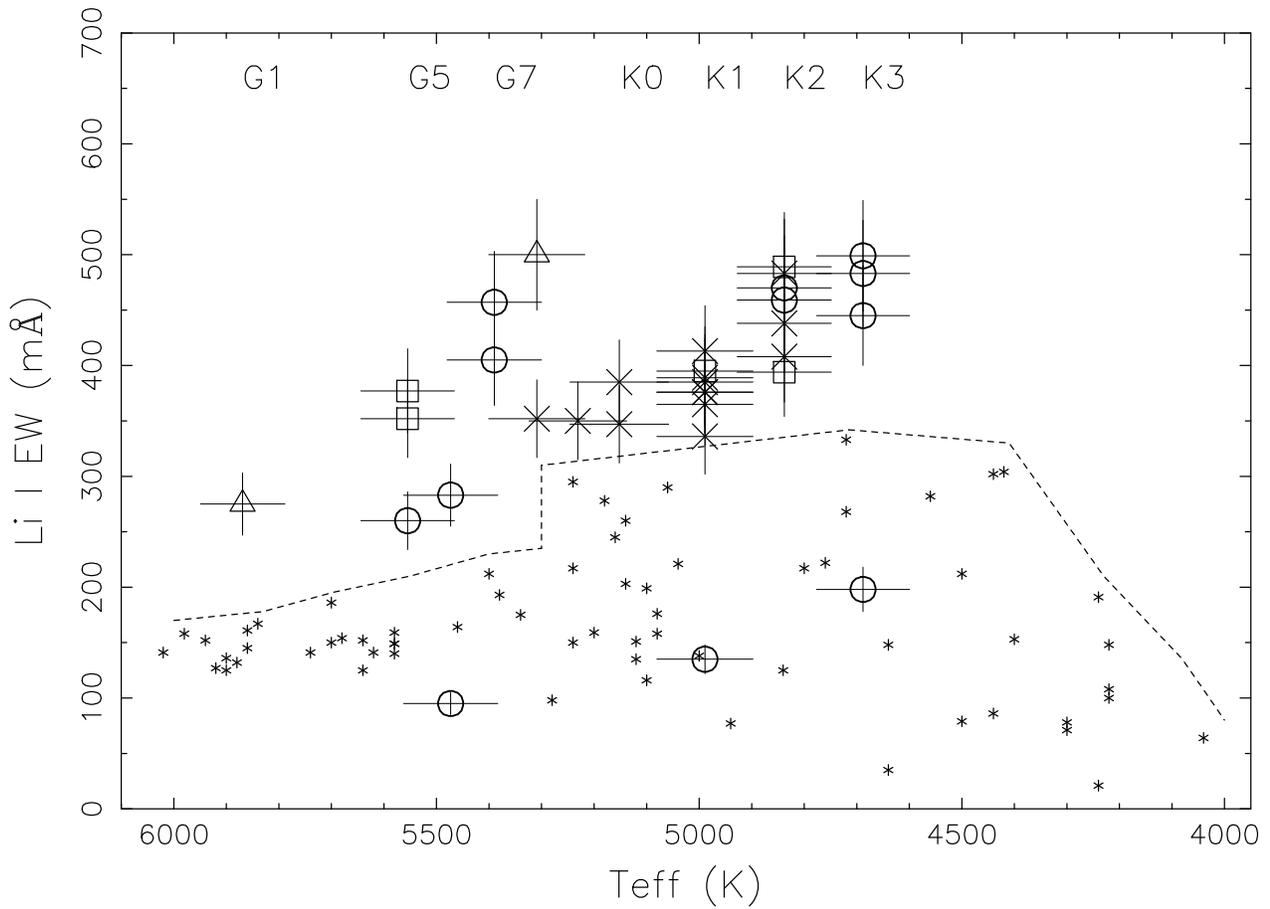}
      \vspace*{5mm}
      \caption{The distribution of Li {\sc i} 6707.8 \r{A} equivalent 
            width versus stellar effective temperature is presented 
            for the sample of WTTS candidates in our target SFRs, as 
            well as for the 100 Myr Pleiades open cluster. The Pleiades 
            data are represented as small asterisks, and are taken from 
            Soderblom et al. (1993a). The Lupus SFR data are depicted 
            by the central crosses, open circles represent the 
            Chamaeleon data, open triangles represent the two Rho 
            Ophiuchus star measurements, whereas the open squares 
            are for the CrA SFR data. The effective temperatures 
            for the young SFR data sample are derived (and 
            interpolated) from a spectral-type versus temperature 
            analysis, for luminosity class {\sc v} stars, performed 
            by de Jager \& Nieuwenhuijzen (1987). Errors bars on 
            the temperatures are estimated from temperature differences 
            between the luminosity class {\sc iv} and class {\sc v} results. 
            The hand-drawn, dashed line represents the Pleiades upper 
            envelope of lithium equivalent widths as a function of mass.}
         \label{Li-youth}
   \end{figure}

\begin{figure}

   \centering
   \includegraphics[angle=0]{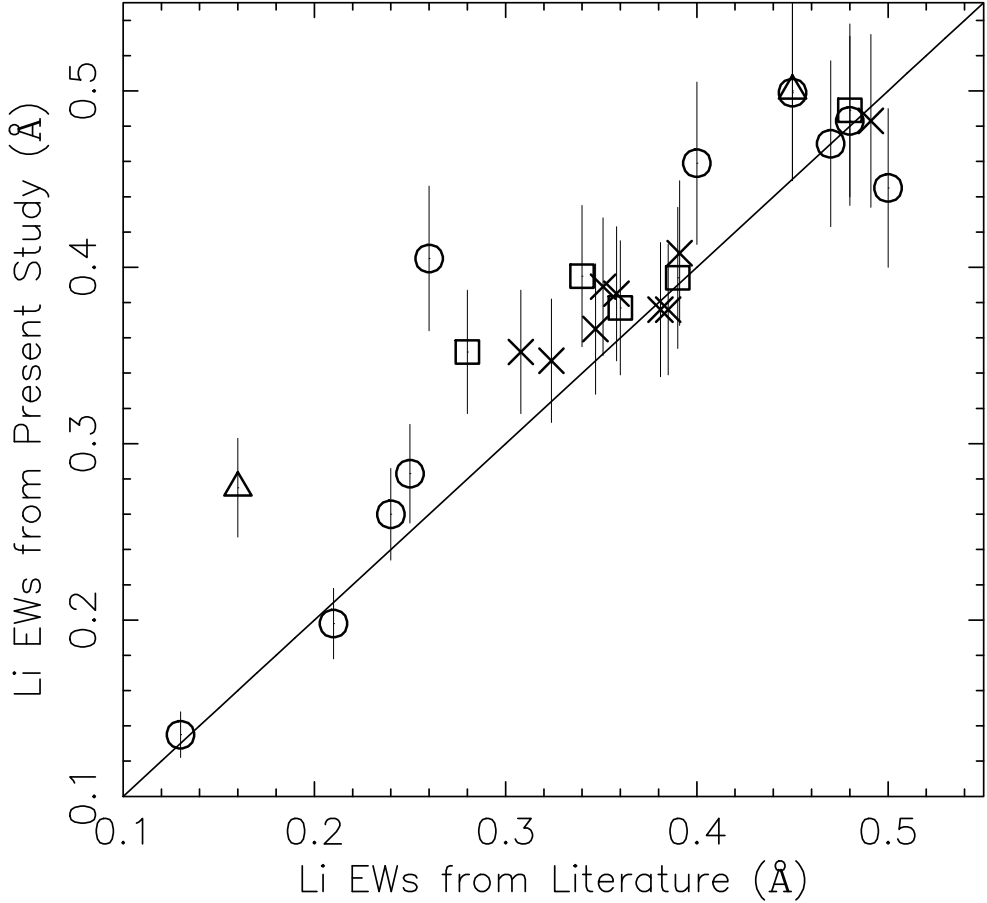}
      \vspace*{5mm}
      \caption{A comparison between the equivalent width measurement 
            scale of the Li 6707.8\r{A} line for the SFRs targets 
            in the current study is presented for the {\sc feros} 
            spectra we have obtained and the extant data in the 
            literature. The symbols for the SFR data source 
            are the same as those presented in Figure~\ref{Li-youth}. 
            The solid line is indicative of equality between 
            our EW measurement system and those found in the 
            literature, and is {\bf not} a fit to the data. We note 
            that an independent data reduction of the {\sc feros} 
            data for RXJ0951.9-7901 yields a lithium EW = 0.28 \r{A}, 
            instead of our $\simeq 0.4$ \r{A} result, far more in 
            line with the value of 0.26 \r{A} to be found in the 
            literature. Such a discrepancy could be due to the 
            rapid rotation and/or possible binary nature of this 
            star.} 
         \label{Li-comp}
   \end{figure}

In further assessing the youthfulness for the stars comprising 
the SFR samples presented in Table~1, we compare 
the adopted lithium EWs for each target, as are detailed 
in Table~3, with similar spectral-type members 
of the 100-Myr Pleiades open cluster. The results of such 
a comparison are presented in Figure~\ref{Li-youth}. For an 
effective temperature scale, we now use the spectral 
type-temperature relationships described in de Jager \& 
Nieuwenhuijzen (1987) because they published temperature 
data for both dwarf and giant luminosity classes, thus 
permitting an estimate of temperature errors. One should 
further note that the \lii EWs for the {\sc feros} observations 
have {\em not} been corrected for the troublesome and 
contaminating Fe {\sc i} line in the blue wing of the \lii 
line, at 6707.44\r{A}. This should be of no great import 
to our results, since for a K2 star, its magnitude is 
of order 15 m\r{A} and is smaller than the measurement 
error of the main \lii line for the majority of our targets. 
Serving as an internal verification, and to reassure ourselves 
of our line measurement quality, we compare the lithium EWs 
we have derived with those of other authors for any {\sc feros} 
target stars in common between our sample and those to be 
found in the extant literature. Such a comparison is 
graphically presented in Figure~\ref{Li-comp}. For all except 
two or three data points, the agreement between our equivalent 
width system and those to be found in the literature is really 
rather good ({\sc rms}$\simeq 34$m\r{A}). There are two objects 
with really quite discordant values, namely, the Rho Ophiuchus 
star RX J1627.1-2419 and the Chamaeleon star RX J0951.9-7901 
(the latter is a suspected SB2 system which may be a mitigating 
factor). 

All targets in the {\sc feros} sample, with the exception 
of the early-type star in the Chamaeleon sample $-$ 
RX J1303.3-7701, contain a clearly defined, and in most 
cases substantial, \lii 6707.8\r{A} line in their 
spectra. However, the most striking feature of the 
Li data presented in Figure~\ref{Li-youth} is that the 
majority of the target stars have lithium EWs considerably 
higher than their similar-mass (\ie similar effective 
temperature) Pleiades counterparts. In fact, their data 
points appear to lie higher even than the upper envelope to 
the Pleiades Li-mass distribution. Lacking a precise 
effective temperature scale for WTTSs, as well as proper 
atmospheric models, we do not attempt to derive Li abundances 
from the measured Li EWs. Nevertheless, that most WTTSs of 
our sample have larger Li EWs than Pleiades late-type dwarfs 
of similar temperature indicates that they are probably 
considerably younger than the Pleiades, and thus qualify as 
young, SFR candidate members (Mart\'{\i}n, Magazz\`{u} \& Rebolo 
1992; Mart\'{\i}n et al. 1994). If their RVs are in agreement 
with the systemic velocities of each respective SFR, they 
must be considered probable {\em young} members of each SFR, 
and as such, are not background or foreground field stars 
rambling through the Galaxy. 

There are however three exceptions. The Chamaeleon {\sc i} \& 
{\sc ii} stars RX J1307.3-7708, RX J1233.5-7523 \& RX J1140.3-8321 
exhibit lithium EWs comparable, or even below, their 
similar-spectral-type counterparts in the 100-Myr Pleiades 
cluster. Moreover, within the error bars, the latter two 
stars have {\sc feros} Li EWs in agreement with previously 
published values.

Understanding the case of RXJ1307.3-7708 is facile. The 
star is an RV non-member and its {\sc 2mass} JHK$^{'}$ 
magnitudes are far too dim for it to be an unobscured WTTS 
member of the Chamaeleon SFR. This object is most likely a 
non-member of the Chamaeleon SFR. The status of the other 
two low-lithium stars is less obvious. Both are RV members 
and have {\sc 2mass} JHK magnitudes which are comparable 
to similar spectral-type stars in the SFR (although, 
see below). Using lithium content for RX J1233.5-7523 \& 
RX J1140.3-8321 as an indicator of extreme youth (compared 
to the Pleiades), with \lii EWs of 135 \& 198 m\r{A} respectively, 
it becomes clear from Figure~\ref{Li-youth} that these stars 
would have to exhibit effective temperatures of $>6000$ K 
to satisfy the criterion of being substantially younger 
than the Pleiades. With published spectral-types of 
K1 and K3/4, such is not the case. These stars are 
potentially older stars than the Pleiades. 

However, as we have already seen, their placement on an 
HRD yields PMS isochronal ages for these stars of 0.5 
and 2.0 Myr respectively. Their absolute status is for 
the present time uncertain, however we retain them as 
photometric, kinematic and isochronal members of the 
Chamaeleon complex, albeit being apparently low-Li stars. 
Both stars are included in the metallicity analysis, 
see \S~\ref{metal} and Table~\ref{tab:parameters}, yielding 
values similar to the remainder of the Chamaeleon sample, 
further hinting at their {\em bona fide} SFR membership.

\subsection{Infra-red Colour and Balmer Emission:}

Some discrimination between a SFR's WTTSs and CTTTs can be probed, 
by plotting the candidates' {\sc 2mass} JHK$^{'}$ data in the form 
of a two-colour J$-$H vs H$-$K$^{'}$ diagram. Those candidates 
with clear infra-red excesses can thus be flagged as likely CTTS 
candidate members of their SFR. 

\begin{figure}
   \centering
   \includegraphics[angle=-90,width=\textwidth]{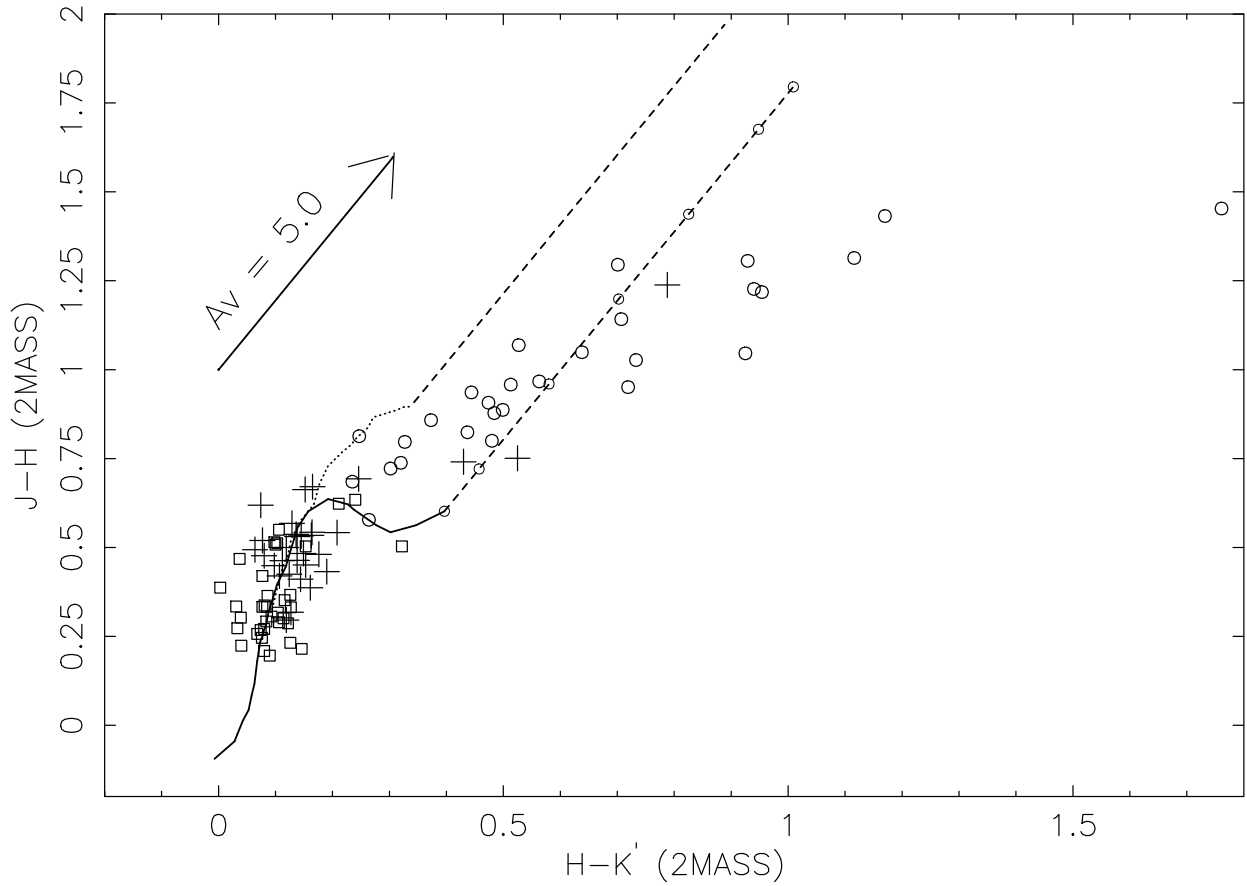}
      \caption{A {\sc 2mass} J$-$H/H$-$K$^{'}$ two-colour diagram 
           is shown for the WTTS sample (crosses) detailed in 
           Table~1. Also plotted are {\sc 2mass} data 
           for a sample of single CTTSs (open circles) located 
           in Orion (identifications taken from Neuh\"{a}user 
           et al. 1995) and a sub-sample of the single, 
           F$\rightarrow$M, main sequence, field stars (open 
           squares) identified by Nidever et al. (2002). The solid 
           line represents a J-H/H-K' relation for the 
           intrinsic colours of field dwarfs as detailed by 
           Bessel \& Brett (1988), transformed onto the {\sc 2mass} 
           JHK$^{'}$ system, whereas the dotted line is the equivalent 
           one for giant stars. The two dashed lines represent reddening 
           vectors originating from the extrema of the Bessel and Brett 
           dwarf and giant sequences.}
         \label{JHK}
   \end{figure}

By exploiting the {\sc 2mass} data for each star in each SFR, as are 
detailed in Table~1, we have investigated just such an 
effect in order to convince ourselves that each sample of young stars 
that we have observed spectroscopically is indeed a {\em bona fide} set 
of WTTSs, and not of the CTTS class. A two colour infra-red diagram 
using data for our WTTS candidates is plotted in Figure~\ref{JHK}. Also 
shown are similar data for a sample of single, classical T-Tauri stars 
in Orion, as identified by Neuh\"{a}user et al. (1995), and a sample 
of nearby, F$\rightarrow$M, main sequence, field stars identified by 
Nidever et al. (2002). All data are taken from the {\sc 2mass} 
all-sky release catalogue of March 2003. We include for completeness 
a J-H/H-K' relation for the intrinsic colours of field dwarfs and giants 
as detailed by Bessel \& Brett (1988), transformed onto the {\sc 2mass} 
JHK$^{'}$ system (Carpenter 2001).

Reassuringly, all bar three of the WTTS candidates of these SFRs 
lie close to the field-star population in J$-$H/H$-$K$^{'}$ space, 
indicating that most of these young, T-Tauri stars are indeed of 
the {\em weak-lined} class and not of the {\em classical} one. The three 
exceptions are for the Chamaeleon {\sc i} star RX J1112.7-7637 and 
the two Rho Ophiuchus candidates. The {\sc 2mass} data for these stars 
indicate that they are more reddened objects or have infra-red 
excesses higher than those seen for any of the other WTTS candidates 
in our sample, with colours more representative of the Orion CTTS 
distribution. In fact, the Rho Ophiuchus star RX J1627.1-2419 appears 
to be highly reddened, at the Av$\simeq 5$ level. As such, we cannot 
have faith in the photospheric purity of this Chamaeleon spectrum and 
the two Rho Ophiuchus ones, in lieu of the fact that their spectra may 
have contributions from a circumstellar component or be suffering from 
higher-than expected reddening. We shall therefore not consider them 
further in any of our analyses, including the metallicity one. It is 
interesting to note in passing, that the \rot value we obtain for the 
apparently extincted Chamaeleon star is about three times faster than 
the value published in the literature. 

One could also argue the point that there are three or four other 
target stars in our sample showing some very small level of 
infra-red discrepancies ($\simeq 0.1$ magnitudes). Their {\sc 2mass} 
data place them just above the shoulder of the Bessel \& Brett 
intrinsic colour curve for dwarfs, and are potentially situated 
along the Bessell \& Brett intrinsic giant-star sequence. These stars 
are identified as the Chamaeleon objects RX J1129.2-7546, RX 
J1140.3-8321, RX J1158.5-7754a \& RX J1159.7-7601. The one star 
in this group whose JHK$^{'}$ data place it furthest away from 
the Bessel \& Brett dwarf curve, some 0.1 magnitudes in H-K$^{'}$ 
colour, is RX J1140.3-8321 $-$ an RV member of the association; 
It is also one of the low-lithium stars identified in 
\S~\ref{lithium}.

In order to be sure that some CTTSs are not contaminating 
our metallicity sample, we have also measured the residual 
H$\alpha$ emission in each of the targets. A perusal of 
the direct integration EW data for the H$\alpha$ line, in 
Table~3, shows that only one object has an 
H$\alpha$ EW $>2.5$\r{A}, RX J1625.6-2613, considerably below 
the canonical limit of 10\r{A} for a star to be classified 
as a {\em classical} type T-Tauri star. This Rho Ophiuchus star 
is already flagged as a possible CTTS (or very reddened 
object), as inferred from its position in the two-colour 
J-H/H-K$^{'}$ diagram. Simply stated, in terms of H$\alpha$ 
emission, only this star does not appear to be of the WTTS 
class.

\subsection{Magnetic Activity and Rotation:}

\begin{figure}
   \centering
   \includegraphics[angle=-90,width=\textwidth]{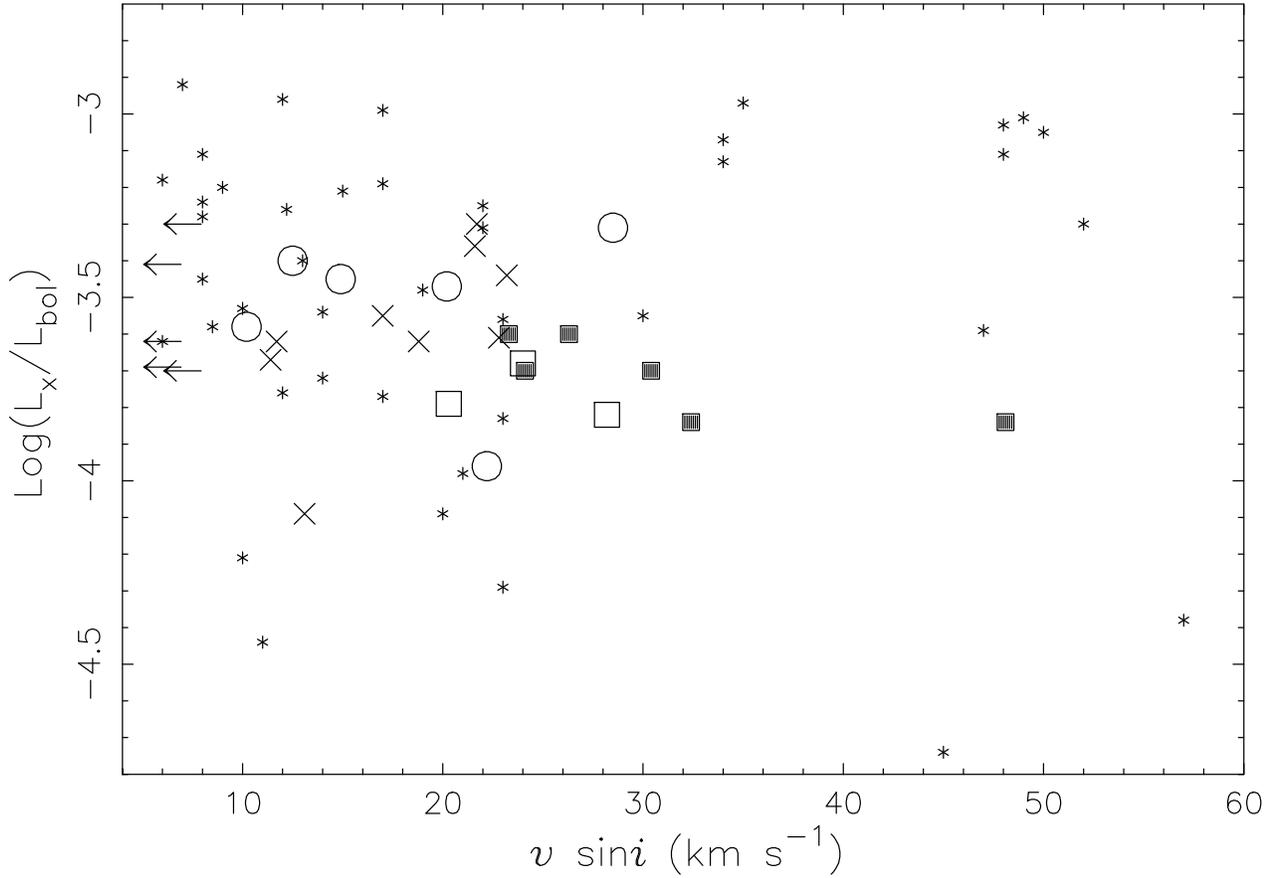}
      \caption{X-ray luminosity, as a fraction of bolometric 
            luminosity, is plotted against spectroscopic 
            rotation rate for SFR candidate-members detailed 
            in Table~2. The symbols for the SFR 
            data source are the same as those presented in 
            Figure~\ref{Li-youth}. X-ray activity-rotation 
            data for the SB2 stars discovered in the {\sc feros} 
            sample are depicted by filled squares, and given 
            the lack of further information, we have allocated 
            $50\%$ of the X-ray luminosity to each component 
            of each SB2. Also plotted for comparison, are the 
            relevant data for a sample of single, solar-type 
            stars in the young (30-50 Myr) IC 2391 \& 2602 
            clusters (small asterisks - Stauffer et al. 1997).}
         \label{Lx-Vsini}
   \end{figure}

The X-ray activity for our WTTS sample is shown graphically in 
Figure~\ref{Lx-Vsini}, in which X-ray luminosity, as a fraction 
of bolometric luminosity, is plotted against spectroscopic 
rotation rate. To enable an estimate and comparison of the 
X-ray activity levels, similar data are also plotted for 
single, solar type members of the 30-50 Myr IC 2602 \& IC 2391 
clusters. Given that the WTTSs exhibit Lx/Lbol values 
comparable, as a function of rotation rate, to the older 
clusters, we assume that this X-ray activity is coronal in 
nature. 

It is clear that members of the IC clusters have attained 
maximal Lx/Lbol values, approaching a plateau or 
{\em saturation}-like level of $10^{-3}$, as is observed for 
solar-type members of main sequence clusters such as the 
Alpha Per and Pleiades (Randich et al. 1996; Stauffer et 
al. 1994). This {\em saturation} in X-ray level is independent 
of rotation rate. However, while we ascertain that the 
target WTTSs are indeed extremely X-ray active, their 
Lx/Lbol values appear to be at a level $\simeq 2-3$ times 
weaker than their IC 2391/2602 counterparts. This maximal 
level of Lx/Lbol emission among the WTTSs appears to be 
plateau-like, and moreover appears independent of rotational 
velocity. Furthermore, for \rot$<30$ km~s$^{-1}$, there 
is a quite considerable dispersion in X-ray emission 
among the WTTSs, at the 0.5 dex level.

The case for analysis of the \ha emission is similar to 
the X-ray one. A photospheric contribution has already 
been subtracted from each WTTS, allowing us the freedom 
to discuss a true chromospheric contribution to the 
line, at least for those stars where we are sure there 
is little or no {\em classical}-like circumstellar 
contribution. A comparison to the older Pleiades cluster 
is possible because Soderblom et al. (1993b) have published 
\ha EWs (and rotation rates) for solar type stars, determined 
in an identical manner to the one presented herein; that 
is to say, where a photospheric \ha profile has already been 
subtracted on a star-by-star basis for each Pleiad. 

\begin{figure}

   \centering
   \includegraphics[angle=-90,width=\textwidth]{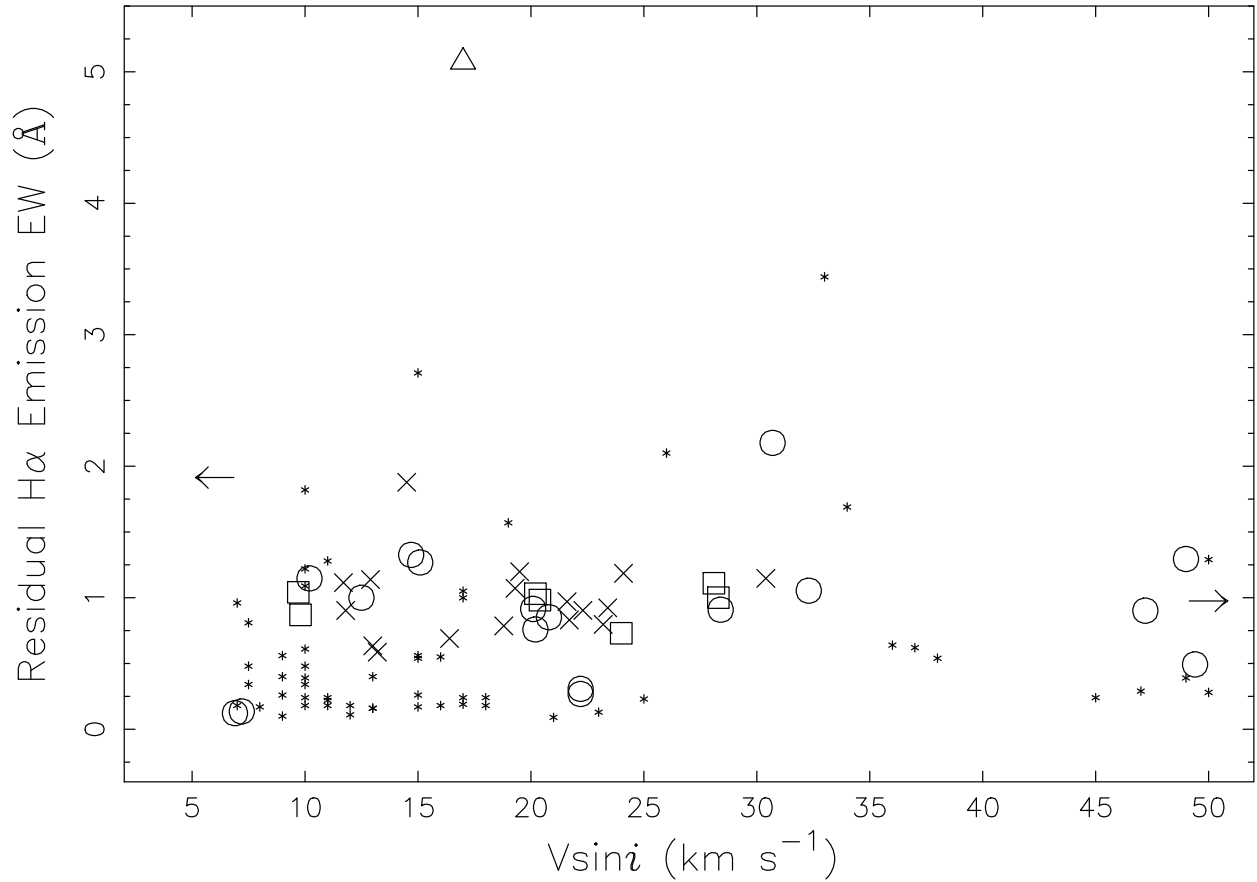}
      \vspace*{5mm}
      \caption{The relationship between residual H$\alpha$ emission strength 
            (direct integration) and projected equatorial velocity for the 
            WTTS candidate members of the target SFRs is presented. The 
            symbols for the SFR data source are the same as those 
            detailed in Figure~\ref{Li-youth}. Also included are data for single, 
            solar-type Pleiads (asterisks - data taken from Soderblom et al. 1993b).}
         \label{Halpha-ROT}
   \end{figure}

A graphical representation of such data is presented in 
Figure~\ref{Halpha-ROT}, where residual emission \ha EWs 
are plotted against spectroscopic rotation rates for the 
{\sc feros} WTTSs and their Pleiad solar-type counterparts. 
There are three striking features; ({\bf a}) The WTTSs are 
indeed chromospherically active at a similar, or indeed 
higher, level than the 100 Myr Pleiades stars.  ({\bf b}) 
As is the case for X-ray emission, there appears to be 
an emission plateau, albeit less well defined than the 
X-ray one, at about $1.5-2.0$ \r{A}, which is more or 
less independent of rotation rate.  ({\bf c}) At any 
given rotation rate, there is considerable scatter 
about the {\em mean} residual H$\alpha$ emission 
level.

\section{Metallicities}\label{metal}

\subsection{Stellar parameters and chemical analysis}

Using our {\sc feros} spectra, we have derived stellar parameters 
and metallicities using the same methodology used by Santos, Israelian 
\& Mayor (2004). In the first instance, EWs for a list of 39 Fe {\sc i} 
and 16 Fe {\sc ii} lines were measured using a Gaussian fitting procedure 
within the IRAF {\sc splot} task. The line-list used is an upgrade 
of the list presented in Santos et al. (2004), with the addition 
of 4 more Fe {\sc ii} lines taken from the literature (see 
Table~5). As before, the $\log{gf}$ values for the 
added lines were computed from an inverted solar analysis using solar 
EWs measured from the Kurucz Solar Atlas (Kurucz, Furenlid \& Brault 
1984), and a Kurucz grid model for the Sun having 
[T$_{\mathrm{eff}}$, $\log{g}$, $\xi_{\mathrm{t}}$, $log{\epsilon}_{Fe}$]
= 5777K,~ 4.44\,dex,~ 1.00\,km~s$^{-1}$,~ 7.47\,dex.

The spectroscopic analysis was done in LTE using the 2002 version 
of the code {\sc moog} (Sneden 1973)\footnote{The code {\sc moog2002} can 
be downloaded at http://verdi.as.utexas.edu/moog.html}, and a grid 
of Kurucz {\sc atlas9} atmospheres (Kurucz 1993). The atmospheric 
parameters for our program stars were obtained from the Fe {\sc i} 
and Fe {\sc ii} lines by iterating until the correlation coefficients 
between $\log{\epsilon}$(Fe {\sc i}) and $\chi_l$, and between 
$\log{\epsilon}$(Fe {\sc i}) and  $\log{({W}_\lambda/\lambda)}$ 
were zero, and the mean abundance given by Fe {\sc i} and Fe 
{\sc ii} lines were the same. This procedure, using iron ionization 
and excitation equilibrium, was shown to give excellent results 
for solar-type dwarfs. The results of our analysis are presented 
in Table~\ref{tab:parameters}, together with the number of Fe 
{\sc i} and Fe {\sc ii} lines used in each case, and the {\sc rms} 
for each set of lines.

\begin{figure}
   \centering
   \includegraphics[angle=0]{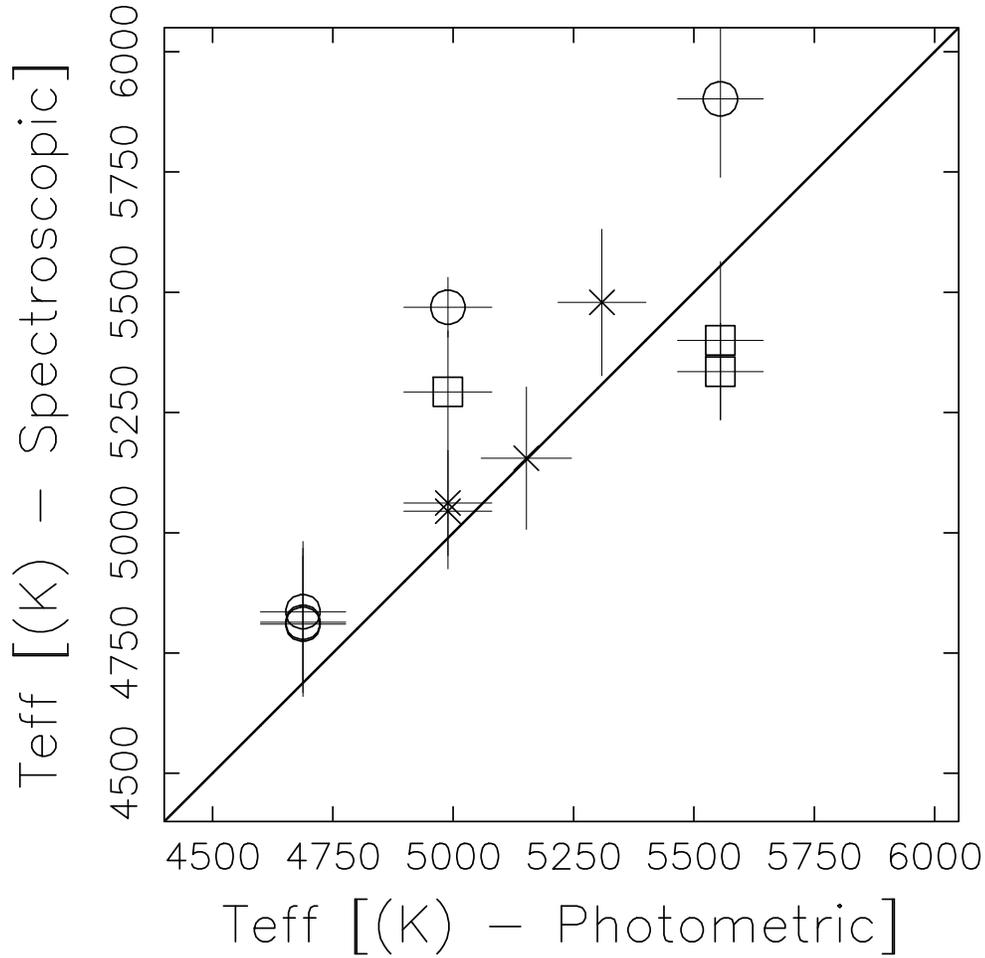}
      \vspace*{5mm}
      \caption{A comparison between the temperature scales derived 
           for the photometric (lithium abundance) analysis and 
           the spectroscopic (metallicity) analysis is presented. 
           The symbols for the SFR data source are the same as 
           those presented in Figure~\ref{Li-youth}. The solid line 
           is indicative of equality between the two temperature 
           systems, and is {\bf not} a fit to the data.}
         \label{Teff-comp}
   \end{figure}

At this point, an interesting digression allows us to compare 
stellar effective temperatures, derived from photometric and 
spectroscopic analyses, for target stars common to lithium 
abundance sample (see \S~\ref{lithium}) and the metallicity one. 
We examine the temperatures derived using spectral type-temperature 
relationships described in de Jager \& Nieuwenhuijzen (1987), and 
for those determined using an Fe-line ionization and excitation 
equilibrium analysis (see Table~\ref{tab:parameters}). The results 
of such a comparison are plotted in Figure~\ref{Teff-comp}. There is 
a general, broad-range agreement ({\sc rms}$\sim 225$ K) between 
the two temperature systems, however, there are also three or four 
targets with quite different photometric and spectroscopic temperatures 
(at the $>300$K level). It is possible that the spectral types for 
these stars, the Chamaeleon objects RXJ 1201.7-7859 \& RXJ 1233.5-7523 
and the CrA star RXJ 1839.0-3726,  are incorrect by a sub-class or two, 
and yet we find no evidence for excessive photometric reddening in 
these stars nor are their lithium EWs low (except for RX J1233.5-7523) 
nor H$\alpha$ emission levels particularly high.

The errors on the atmospheric parameters were derived as in the same 
manner as that presented in Gonzalez \& Vanture (1998). Since the 
{\sc rms} around the average abundance given by the Fe {\sc i} lines 
is used to compute the final uncertainties in the [Fe/H] abundance 
(instead of the {\sc rms}/$\sqrt{n}$, with $n$ being the number of 
lines used), it may be that these errors are slightly over-estimated.

The relatively low S/N ratio of our spectra (always below $\sim$100), 
together with the fact that most of our targets are cool K-dwarfs 
(with small Fe {\sc ii} lines), makes our metallicity analysis difficult 
and in a few cases impossible. In all, we could only obtain satisfactory 
stellar parameters for 12 WTTSs in our entire sample. Furthermore, 
only for a very few stars rotating faster than $\sim$15 km~s$^{-1}$ could 
we measure the line EWs, although these include most of the initial 
targets. In these cases, however, the results must be viewed with caution, 
as line-blending for rapid rotators can severely limit the precision of 
the EW measurements and hence the metallicity results.

The micro-turbulence velocities [V$_{\small t}$] we have derived are 
detailed in Table~\ref{tab:parameters}, and are considerably above 
the ones found for main-sequence dwarfs with similar temperature 
(see \eg Santos et al. 2004). This may be due to the effect of 
magnetic fields, as discussed in Steenbock \& Holweger (1981 - 
however, see also discussion in Padgett 1996).

Recently, Morel et al. (2004) have studied the detailed effects 
of stellar spots affecting the determination of stellar parameters 
and metallicity for RS CVn stars. Their results show that the final 
abundances are indeed affected, although not always strongly, by 
the presence of photospheric spots. This issue may thus have 
particular importance when studying young, magnetically stars like 
the ones in our sample. These effects may in part explain the 
metallicity dispersion observed in stars belonging to the same SFR 
(see also next section). Unfortunately, the quality of our spectra 
and the relatively large error bars we calculate for our derived [Fe/H] 
results hinders any clear and productive discussion of the effect of 
photospheric inhomogeneities on metallicity determinations.

Finally, in our analysis we have considered that our target stars 
do not have any significant spectroscopic veiling. At this stage, 
this is a reasonable supposition because as the {\sc 2mass} photometry 
data indicate, our data ensemble in each SFR is dominated by 
stars of the {\em weak-lined} class (Hartigan, Edwards \& Ghandour 1995). 
Small amounts of low-level veiling could however be responsible for 
some representation of the dispersion in the derived [Fe/H] values.

\subsection{The metallicity of the Star Formation Regions}

In total we have obtained metallicity determinations for 12 
WTTS stars in 3 different SFRs: Five objects in Chamaeleon 
(Cha), four in Lupus (Lup), and three in Corona Australis (CrA).

In Table~7 we present the average and {\sc rms} of the 
metallicity values computed, considering all the stars with [Fe/H] 
values measured in each SFR, as well as taking only those stars with 
projected rotational velocity values $v\sin{i}<15$km~s$^{-1}$. As 
mentioned above, stars with larger $v\sin{i}$ values will have higher 
uncertainties for their derived stellar parameters and metallicities; 
these indeed usually present the largest deviations from the mean 
[Fe/H] (see Table~\ref{tab:parameters}).

In all cases, the average metallicity obtained for each SFR is below 
solar, typically in the range 0.0 to $-$0.2\,dex. This is in agreement 
with the results from X-ray studies of SFRs (e.g., Pallavicini, 
Franciosini \& Randich 2004). X-ray observations, however, sample 
the stellar corona, where the abundances do not reflect the real 
photospheric abundances (see \eg review by Audard 2004). Given 
that the peak in the metallicity distribution of stars in the solar 
neighbourhood is also below solar (\eg Allende Prieto et al. 2004; 
Taylor \& Croxall 2005; Santos et al. 2005), the metallicities we 
derive for these relatively nearby SFRs maybe not be such a surprise.

Globally, the relatively small dispersion in the [Fe/H] values for a 
given SFR attest to the quality of our measurements. The results are 
clearly below the derived error bars in the individual [Fe/H] estimates, 
suggesting that these latter may be overestimated (see previous section). 
Also, the small dispersion gives us confidence that the use of the 
sample WTTSs we have targeted may be indeed a good indicator of the 
metallicity for each of the study regions.

Padgett (1996) have also derived stellar metallicities for stars in the 
Cha SFR. The average metallicity they have derived for their five objects 
is $-0.06 \pm 0.14$, a value that is compatible with the CrA results listed 
in Table~7.

\section{Summary}

Using high-resolution spectra of X-ray selected, WTTS candidate members 
of the Lupus, Chamaeleon {\sc i} \& {\sc ii} and CrA star forming regions, 
we have established kinematic membership of their parent associations, 
proved that these stars are young, confirmed that each is of the {\em weak-line} 
class of T-Tauri star and is magnetically active. These analyses have 
allowed us to create a sample of high quality spectra for youthful, 
single {\em bona fide} members of each region. 

An analysis of radial velocities derived for WTTS candidate members 
in each region shows that majority of the samples are 1-d kinematic 
members of their parent associations. We find a mean RV for objects 
in the Lupus cloud to be ${\overline {RV}}=+2.6 \pm 1.8$ km~s$^{-1}$, 
for the Chamaeleon {\sc i} \& {\sc ii} clouds, 
${\overline {RV}}=+12.8 \pm 3.6$ km~s$^{-1}$ whereas for the CrA 
SFR, we find ${\overline {RV}}=-1.1 \pm 0.5$ km~s$^{-1}$, consistent 
with earlier studies in these regions.  

Using extant photometry and distance estimates, we have determined 
fundamental properties for SFR candidate members such as luminosities 
and effective temperatures. These parameters, used in concert with 
pre-main sequence theoretical stellar models, allow us to determine 
model-dependent isochronal ages of the Lupus, Chamaeleon and CrA 
targets to be $9.1 \pm 2.1$ Myr, $4.5 \pm 1.6$ Myr and $9.0 \pm 3.9$ 
Myr respectively.

The youthfulness of the stars is further confirmed by measuring Li 
{\sc i} 6707.8 \r{A} equivalent widths and comparing them to similar data 
for single, solar-type members of the 100 Myr Pleiades open cluster. 
All bar three target stars have Li EWs at or above the upper envelope 
of the lithium EW distribution, as a function of mass, in the Pleiades. 

By comparing coronal and chromospheric activity indicators to solar-type 
stars in the young IC 2602/2391 (30-50 Myr) and/or Pleiades clusters, 
we find that the majority of our SFR targets are extremely magnetically 
active with L$_{x}$/L$_{bol}$ values and H$\alpha$ EWs which are almost, 
but not quite, as active as their maximally-active older, solar-type 
counterparts. 

The better quality spectra of single, genuine members of each SFR are 
used to investigate the primary scientific question motivating this 
project. Are any of these star formation regions metal-rich, and can 
we detect a metallicity spread among their members ? Such project 
goals are crucial to our future studies of these regions in our efforts 
to detect and then characterize any extra-solar planets orbiting these 
young stars. To this end, for the higher quality spectra, we have 
performed an iron-line metallicity analysis for five (5) stars in 
Chamaeleon, four (4) stars in Lupus and three (3) stars in the CrA 
SFRs. All three regions are actually slightly metal-poor, with 
marginally sub-solar metallicities, with $<$[Fe/H]$> = -0.11 \pm 0.14$, 
$-0.10 \pm 0.04$ \& $-0.04 \pm 0.05$  respectively.

\begin{acknowledgements}

Most of this project was conducted within the framework of the 
European Research Training Network entitled {\em The Formation and 
Evolution of Young Stellar Clusters} (HPRN-CT-2000-00155), from 
which we gratefully acknowledge support. Some support has also 
been derived from an {\sc nsf} {\em career award} grant $\# 0349075$, 
which is also gratefully acknowledged. Support from Funda\c{c}\~ao 
para a Ci\^encia e a Tecnologia (Portugal) to N.C.S. in the form 
of a scholarship (SFRH/BPD/8116/2002) and grant reference 
POCI/CTE-AST/56453/2004 is gratefully acknowledged. Our thanks 
and gratitude go to the staff at the European Southern Observatory, 
La Silla, Chile. Some sections of this project were completed during 
a research visit by N. Santos to the European Southern Observatory, 
supported by its Visitor Scientist Programme.

This research has made use of the {\sc simbad} database, operated 
at the Centre de Donn\'{e}es astronomiques de Strasbourg, La 
France, and the Leicester Database and Archive Service at the 
Department of Physics and Astronomy, Leicester University, 
UK. The authors have been Guest Users of the Canadian Astronomy 
Data Centre, which is operated by the Herzberg Institute of 
Astrophysics, National Research Council of Canada. This 
publication makes use of data products from the Two Micron 
All Sky Survey, which is a joint project of the University 
of Massachusetts and the Infrared Processing and Analysis 
Center/California Institute of Technology, funded by the 
National Aeronautics and Space Administration and the 
National Science Foundation. This research has also made 
use of {\sc nasa}'s Astrophysics Data System. 

DJJ would also like to thank Mrs Judith Pryer for much 
love and support during this work, and her continued 
positive influences. Je veux remercier ma petite 
be-draggled hag, Jude, pour son amour et soutien. 
Elle est l'etoile la plus brilliante dans ma vie. 
DJJ gratefully acknowledges the hospitality of 
Alan \& Irene Constable, of Keivan Stassun \& Gina 
Brissenden, and of Sara \& S{\o}ren Meibom, whose 
warmth, kindness and homes were selflessly and freely 
given during several research visits to various 
institutions. 

\end{acknowledgements}


%
%
\newpage

\newpage

\newpage

\newpage

\newpage

\newpage

\newpage 

\newpage

%
%

%
%

\newpage

%
%

%
%
   \begin{table}
      \caption[]{Basic astrometric/photometric data for the WTTS samples, observed 
             spectroscopically, in each SFR .}
         \label{inputcat}
\begin{tabular}{llccrccccc}
\hline\hline
TARGET$^{\mathrm{a}}$ & Alt. ID &  RA(2000)$^{\mathrm{b}}$ & DEC(2000)$^{\mathrm{b}}$ & V~~~ & V$-$I & Sp. Ty. & J$^{\mathrm{c}}$ ($\pm$ err) & 
H$^{\mathrm{c}}$ ($\pm$ err) & K$^{'\mathrm{c}}$ ($\pm$ err) \\ 
\hline
Lupus : & & & & & & & & &  \\ \hline
RX J1507.2-3505  & K 3        & 15 07 14.6 & -35 04 58.9 & 10.53 & 1.26 & K0 & $8.893 \pm 0.021$ & $8.416 \pm 0.053$ & $8.336 \pm 0.023$ \\
RX J1507.4-4601  & K 4        & 15 07 27.6 & -46 01 06.0 & 11.71 &      & K1 & $9.735 \pm 0.024$ & $9.284 \pm 0.030$ & $9.131 \pm 0.021$ \\
RX J1515.7-3331  & K 20       & 15 15 45.4 & -33 31 59.1 & 10.69 & 1.00 & K0 & $8.981 \pm 0.022$ & $8.461 \pm 0.020$ & $8.384 \pm 0.023$ \\
RX J1518.4-3738  & K 25       & 15 18 26.9 & -37 38 02.5 & 10.92 & 1.01 & K1 & $9.081 \pm 0.035$ & $8.618 \pm 0.045$ & $8.506 \pm 0.021$ \\
 (V* LW Lup) \\
RX J1523.4-4055  & K 29       & 15 23 25.6 & -40 55 46.8 & 11.87 & 1.16 & K2 & $9.957 \pm 0.023$ & $9.389 \pm 0.023$ & $9.260 \pm 0.023$ \\
 (V* MM Lup) \\ \hline
RX J1524.5-3652  & K 32       & 15 24 32.4 & -36 52 03.3 & 11.30 & 1.01 & K1 & $9.549 \pm 0.023$ & $9.049 \pm 0.021$ & $8.930 \pm 0.019$ \\
  (V* MP Lup) \\
RX J1525.0-3604  & K 33       & 15 25 03.6 & -36 04 45.8 & 10.92 & 1.09 & K1 & $8.998 \pm 0.030$ & $8.465 \pm 0.038$ & $8.320 \pm 0.022$ \\
RX J1529.6-3546  & K 42       & 15 29 38.6 & -35 46 52.1 &       &      & K1 & $8.788 \pm 0.020$ & $8.257 \pm 0.033$ & $8.119 \pm 0.016$ \\
   (PPM 293919) \\
RX J1529.7-3628  & K 43       & 15 29 47.3 & -36 28 38.0 &       &      & K2 & $9.643 \pm 0.025$ & $9.106 \pm 0.027$ & $8.970 \pm 0.025$ \\
RX J1546.6-3618  & K 70       & 15 46 41.0 & -36 18 47.5 & 11.28 & 1.07 & K1 & $9.490 \pm 0.024$ & $8.947 \pm 0.025$ & $8.783 \pm 0.021$ \\
RX J1547.6-4018  & K 73       & 15 47 41.8 & -40 18 26.7 & 11.08 & 1.05 & K1 & $9.294 \pm 0.024$ & $8.811 \pm 0.034$ & $8.662 \pm 0.025$ \\
RX J1601.1-3320  & K 93       & 16 01 08.9 & -33 20 14.1 & 10.88 & 1.07 & G8 & $9.027 \pm 0.029$ & $8.552 \pm 0.034$ &                \\ 
   (V* MZ Lup) \\
RX J1605.7-3905  & K 103      & 16 05 45.1 & -39 06 06.8 & 10.49 & 0.90 & G9 & $8.910 \pm 0.029$ & $8.523 \pm 0.045$ & $8.362 \pm 0.027$ \\
   (PPM 747449) \\
RX J1608.9-3905  & K 116      & 16 08 54.2 & -39 06 06.4 & 10.88 & 1.14 & K2 & $8.909 \pm 0.030$ & $8.375 \pm 0.047$ & $8.212 \pm 0.029$ \\ \hline
Chamaeleon {\sc i} $\&$ {\sc ii}: & & & & & & & & &  \\ \hline
RX J0850.1-7554  &            & 08 50 05.7 & -75 54 38.9 & 10.57 & 0.82 & G6 & $9.259 \pm 0.026$ & $8.848 \pm 0.025$ & $8.704 \pm 0.019$ \\
RX J0951.9-7901  &            & 09 51 51.1 & -79 01 38.8 & 10.22 & 0.93 & G7 & $8.587 \pm 0.032$ & $8.138 \pm 0.034$ & $8.040 \pm 0.029$ \\
(HD 86356) \\
RX J1112.7-7637  & HBC 588  & 11 12 25.0 & -76 37 06.2 & 12.43 & 1.41 & K0/4 & $9.275 \pm 0.030$ & $8.524 \pm 0.055$ & $7.999 \pm 0.031$ \\
    (CHXR 50) \\
RX J1129.2-7546  &          & 11 29 13.2 & -75 46 26.2 & 12.91 & 1.76 & K3 & $9.817 \pm 0.026$ & $9.124 \pm 0.021$ & $8.878 \pm 0.024$ \\
RX J1140.3-8321  &          & 11 40 17.5 & -83 21 00.8 & 11.56 & 1.33 & K3/4 & $9.328 \pm 0.023$ & $8.709 \pm 0.045$ & $8.635 \pm 0.019$ \\ 
RX J1158.5-7754a &          & 11 58 28.7 & -77 54 29.2 & 10.59 & 1.43 & K2/4 & $8.219 \pm 0.029$ & $7.556 \pm 0.042$ & $7.404 \pm 0.021$ \\
RX J1159.7-7601  &          & 11 59 42.9 & -76 01 26.0 & 11.30 & 1.37 & K2/4 & $9.140 \pm 0.027$ & $8.469 \pm 0.038$ & $8.304 \pm 0.027$ \\
RX J1201.7-7859  &          & 12 01 39.7 & -78 59 16.6 & 8.59  &      & G5 & $7.263 \pm 0.027$ & $6.967 \pm 0.044$ & $6.848 \pm 0.018$ \\ 
 (HD 104467) \\ 
RX J1233.5-7523  &          & 12 33 30.6 & -75 23 11.5 & 9.47  &      & K1 & $8.201 \pm 0.020$ & $7.883 \pm 0.040$ & $7.756 \pm 0.040$ \\
(HD 109138) \\
RX J1239.4-7502  &          & 12 39 21.8 & -75 02 38.9 & 10.24 & 1.06 & K2/3 & $8.434 \pm 0.021$ & $7.953 \pm 0.033$ & $7.777 \pm 0.021$ \\ 
RX J1303.5-7701  &          & 13 03 31.3 & -77 01 53.7 &       &      & G7 &                &                &                \\ 
(CHIIXR24) \\
RX J1307.3-7708  &          & 13 07 15.2 & -77 07 51.0 &       &    & G6 & $12.004 \pm 0.023$ & $11.584 \pm 0.024$ & $11.477 \pm 0.025$ \\ 
(CHIIXR35) \\\hline
Rho Ophiuchus : & & & & & & &  \\ \hline
RX J1627.1-2419  &            & 16 27 10.3 & -24 19 12.3 &       &      & G1 & $8.745 \pm 0.027$ & $7.507 \pm 0.038$ & $6.719 \pm 0.024$ \\
RX J1625.6-2613  &            & 16 25 38.5 & -26 13 53.4 &       &      & G8 & $8.688 \pm 0.019$ & $7.947 \pm 0.055$ & $7.517 \pm 0.024$ \\ \hline
CrA SFR : & & & & & & & & &  \\ \hline
CrAPMS  2        & HBC 678    & 19 02 01.9 & -37 07 42.9 & 10.44 & 0.93 & G5 & $8.903 \pm 0.018$ & $8.478 \pm 0.042$ & $8.354 \pm 0.027$ \\ 
(V* V702 CrA)  \\
RX J1839.0-3726  & NWC-13     & 18 39 05.1 & -37 26 20.5 & 10.81 & 1.00 & K1 & $9.075 \pm 0.021$ & $8.581 \pm 0.029$ & $8.517 \pm 0.023$ \\
RX J1853.1-3609  & NWC-55     & 18 53 05.9 & -36 10 22.8 &  9.60 &      & K2 & $7.905 \pm 0.023$ & $7.473 \pm 0.049$ & $7.283 \pm 0.020$ \\
(HD 174656) \\
RX J1917.4-3756  & NWC-135    & 19 17 23.8 & -37 56 49.5 &  9.90 &      & K2 & $8.217 \pm 0.029$ & $7.675 \pm 0.049$ & $7.467 \pm 0.029$ \\
(SAO 211129) \\
CrAPMS  4SE      &            & 18 57 20.7 & -36 43 00.3 & 10.95 & 1.00 & G5 & $9.271 \pm 0.024$ & $8.807 \pm 0.033$ & $8.669 \pm 0.023$ \\ \hline 
\end{tabular} 
\vspace*{1mm}
\begin{list}{}
\item ${\mathrm{a}}~ - $ Nomenclature based on Rosat All-Sky Survey detections. [http://www.xray.mpe.mpg.de/cgi-bin/rosat/rosat-survey]
\vspace*{1mm}
\item ${\mathrm{b}}~ - $ Astrometric data determined from Digitized Sky Survey images. [http://cadcwww.dao.nrc.ca/cadcbin/getdss$\#1$]
\vspace*{1mm}
\item ${\mathrm{c}}~ - $ JHK data taken from the {\sc 2mass} All-Sky Release Point Source catalogue (March 2003). 
[http://pegasus.phast.umass.edu/2mass.html]
\end{list}
\end{table}

\newpage

   \begin{table}
      \caption[]{An X-ray and kinematic data ensemble for each star detailed 
              in Table~\ref{inputcat}.}
         \label{spectro}
\begin{tabular}{llllllllll}
\hline\hline
TARGET           & {\sc rass} X-ray$^{\mathrm{a,b,c}}$ & Log L$^{d}_{x}$       & Log (L$_{x}$/  & P$_{rot}$ & \rot  & HJD  & RV  & \rot & Comments$^{\mathrm{e}}$  \\  
                 & ct rate [ct s$^{-1}$]  &  [erg s$^{-1}$]  & L$_{bol}$) & [days] &  [km~s$^{-1}$] & (+2450000) & [km~s$^{-1}$]     &  [km~s$^{-1}$] &  \\ 
\hline
Lupus :          &                      &        &       &      &         &            &           &      &               \\ \hline
RX J1507.2-3505  & $0.07 \pm 0.02$      & 29.99  & -4.09 &      &  15.0   &  2713.762  &    +2.1   & 13.0 & RV = +1.4     \\
                 &                      &        &       &      &         &  2718.766  &    +2.2   & 13.2 &               \\
RX J1507.4-4601  & $0.06 \pm 0.02$      & 29.93  & -3.30 &      &         &  2718.790  &    +6.9   & 21.7 &               \\
RX J1515.7-3331  & $0.10 \pm 0.02$      & 30.15  & -3.61 & 2.28 &  25.0   &  2713.860  &    +1.1   & 22.3 &               \\
                 &                      &        &       &      &         &  2718.925  &    +1.9   & 23.4 &               \\
RX J1518.4-3738  & $0.13 \pm 0.02$      & 30.26  & -3.36 & 2.98 &  25.0   &  2718.746  &    +3.7   & 21.6 & RV = +3.7     \\ 
RX J1523.4-4055  & $0.037 \pm 0.012$    & 29.71  & -3.62 & 4.15 &   9.9   &  2719.769  &    +5.6   & 11.7 & RV = +6.8     \\
RX J1524.5-3652  & $0.06 \pm 0.02$      & 29.93  & -3.55 & 2.91 &  19.0   &  2713.826  &    +4.4   & 16.4 & RV = +4.2     \\
                 &                      &        &       &      &         &  2719.792  &    +4.6   & 17.6 &               \\
RX J1525.0-3604  & $0.13 \pm 0.02$      & 30.26  & -3.44 &      &         &  2718.852  &    +4.8   & 23.2 &               \\
RX J1529.6-3546  & $0.11 \pm 0.02$      & 30.19  &       &      &         &  2718.879  &   +16.4   & 19.3 & SB2           \\
                 &                      &        &       &      &         &  2718.879  &  $-11.3$  & 19.5 & SB2           \\
RX J1529.7-3628  & $0.06 \pm 0.02$      & 29.93  &       &      &         &  2719.715  &    +3.2   & 14.5 &               \\
RX J1546.6-3618  & $0.05 \pm 0.02$      & 29.85  & -3.69 &      &   7.1   &  2713.705  &    +7.2   & $\leq 6$ & RV = +2.5 \\
                 &                      &        &       &      &         &  2719.747  &    +7.3   & $\leq 6$ &           \\
RX J1547.6-4018  & $0.06 \pm 0.01$      & 29.93  & -3.67 &      &  10.0   &  2713.737  &    +3.1   & 11.8 & RV = +2.8     \\
                 &                      &        &       &      &         &  2718.916  &    +3.2   & 12.9 &               \\
RX J1601.1-3320  & $0.10 \pm 0.02$      & 30.15  & -3.62 & 3.78 &  21.9   &  2713.799  &    +2.3   & 18.8 & RV = +3.0     \\
RX J1605.7-3905  & $0.16 \pm 0.02$      & 30.35  & -3.40 &      &         &  2713.780  &    +2.2   & $> 50$ &             \\
RX J1608.9-3905  &                      &        &       &      & $\sim49$ & 2718.900  &    +9.7   & 30.4 & RV = $-0.6$   \\
                 &                      &        &       &      &         &  2718.900  &  $-25.8$  & 24.1 & SB2           \\ \hline
Chamaeleon {\sc i} $\&$ {\sc ii} &        &        &       &      &         &            &            &      &               \\ \hline
RX J0850.1-7554  & $0.17 \pm 0.02$      & 30.68  & -3.30 &      &  45     &  2712.590  &   +17.1   & 49.4 & RV = +15.5    \\
                 &                      &        &       &      &         &  2718.521  &   +35.0   & 26.5 & SB2?          \\
                 &                      &        &       &      &         &  2718.521  &   $-0.8$  & 23.3 & SB2?          \\ 
RX J0951.9-7901  & $0.17 \pm 0.02$      & 30.68  & -3.54 &      &  73     &  2712.620  &   +39.8   & 49.0 & RV = +12.2    \\
                 &                      &        &       &      &         &  2712.620  &  $-20.3$  & 32.6 & SB2?          \\
                 &                      &        &       &      &         &  2718.533  &   +38.1   & 47.2 & SB2?          \\
                 &                      &        &       &      &         &  2718.533  &  $-35.1$  & 32.3 & SB2?          \\ 
RX J1112.7-7637  & $0.08 \pm 0.02$      & 30.36  & -3.31 &      &  11     &  2712.694  &   +13.6   & 30.7 & RV = +16.0    \\
                 &                      &        &       &      &         &  2718.546  &   +15.2   & 28.4 &               \\ 
RX J1129.2-7546  &                      &        &       &      &  20     &  2713.628  &   +15.9   & 20.8 & RV = +11.4    \\
                 &                      &        &       &      &         &  2719.528  &   +15.5   & 19.9 &               \\
RX J1140.3-8321  & $0.11 \pm 0.02$      & 30.50  & -3.40 &      &  13     &  2718.563  &   +14.0   & 12.5 & RV = +10.5    \\
RX J1158.5-7754a & $0.30 \pm 0.04$      & 30.93  & -3.45 &      &  11     &  2713.656  &   +10.4   & 15.1 & RV = +13.1    \\
                 &                      &        &       &      &         &  2718.614  &   +10.1   & 14.7 &               \\
RX J1159.7-7601  & $0.10 \pm 0.02$      & 30.46  & -3.58 &      &  10     &  2718.636  &   +15.1   & 10.2 & RV = +13.1    \\
RX J1201.7-7859  & $0.20 \pm 0.02$      & 30.76  & -3.96 &      &  21     &  2712.734  &   +12.3   & 22.2 & RV = +10.0    \\
                 &                      &        &       &      &         &  2718.678  &   +13.6   & 22.2 & RV = +22.3    \\
RX J1233.5-7523  &                      &        &       &      &   7     &  2712.760  &   +14.8   &  7.2 & RV = +14.0    \\
                 &                      &        &       &      &         &  2718.716  &   +16.0   &  6.9 & RV = +15.1    \\
RX J1239.4-7502  & $0.19 \pm 0.02$      & 30.72  & -3.47 &      &  19     &  2713.675  &   +14.5   & 20.2 & RV = +13.1    \\
                 &                      &        &       &      &         &  2718.726  &   +14.2   & 20.1 &               \\
RX J1303.5-7701  &                      &        &       &      &         &  2712.779  & $-121.1$  &      & Early-Type    \\
RX J1307.3-7708  &                      &        &       &      &         &  2712.794  &  $-25.3$  & $\leq 6$ &           \\ \hline
\end{tabular}
\vspace*{1mm}
\begin{list}{}
\item ${\mathrm{a}}~$ The standard {\sc rass} conversion factor, for negligible interstellar 
absorption, of $6\times10^{-12}$ erg cm$^{-2}$ count$^{-1}$ has been used.
\item ${\mathrm{b}}~$ These fluxes are lower limits, as it is unlikely that these 
stars represent a sample of unabsorbed X-ray emitters (see text).
\item ${\mathrm{c}}~$ For the ROSAT passband  $0.1-2.4$ keV.
\item ${\mathrm{d}}~$ Distances used for luminosity calculations: Lupus, d=140 pc 
(Hughes et al 1993; Bertout, Robichon \& Arenou 1999; de Zeeuw et al. 1999); 
Chamaeleon, d=160 pc (Bertout et al. 1999; Rebull et al. 2004 and references therein); 
CrA SFR, d=120 pc (Morraco \& Rydgren 1981; Casey et al. 1998).
\item ${\mathrm{e}}~$ Lupus Vsini/Periods/RV data are taken from Wichmann et al. 
(1998, 1999); Chamaeleon Vsini/RV data are taken from Covino et al. (1997), 
Cutispoto et al. (2002) \& Nordstr\"{o}m et al. (2004); CrA SFR Vsini/RVs/Period 
data are taken from Neuh\"{a}user et al. (2000), Franchini et al. (1992), 
Walter et al. (1997) \& Shevchenko et al. (1995).
\end{list}
%
\end{table}

   \begin{table}
\begin{tabular}{llllllllll}
\hline\hline
TARGET           & {\sc rass} X-ray$^{\mathrm{a,b,c}}$ & Log L$^{d}_{x}$       & Log (L$_{x}$/  & P$_{rot}$ & \rot  & HJD  & RV  & \rot & Comments$^{\mathrm{e}}$  \\  
                 & ct rate [ct s$^{-1}$]  &  [erg s$^{-1}$]  & L$_{bol}$) & [days] &  [km~s$^{-1}$] & (+2450000) & [km~s$^{-1}$]     &  [km~s$^{-1}$] &  \\ 
\hline
Rho Ophiuchus :  &                      &        &       &      &         &            &           &      &               \\ \hline
RX J1627.1-2419  &                      &        &       &      &         &  2712.852  &   $-2.7$  &      &               \\
RX J1625.6-2613  &                      &        &       &      &         &  2712.886  &   $-3.5$  & 17.0 &               \\ \hline
CrA SFR :        &                      &        &       &      &         &            &           &      &                   \\ \hline
CrAPMS  2        & $0.08 \pm 0.02$      & 29.93  & -3.79 & 2.79 &  20     &  2713.909  &   $-0.7$  & 20.2 & RV = $-2.2,-1.2$  \\
                 &                      &        &       &      &         &  2719.923  &     0.0   & 20.4 & RV = $-2.6$       \\ 
RX J1839.0-3726  &                      &        &       &      &  17.8   &  2719.869  &   $-1.6$  &  9.8 & RV = $-4.8$       \\
RX J1853.1-3609  & $0.19 \pm 0.04$      & 30.28  & -3.68 &      &         &  2719.837  &   $-3.2$  & 24.0 &                   \\
RX J1917.4-3756  & $0.19 \pm 0.03$      & 30.28  & -3.82 &      &         &  2713.920  &   $-0.9$  & 28.1 &                   \\
                 &                      &        &       &      &         &  2719.817  &   $+0.5$  & 28.3 &                   \\
CrAPMS  4SE      &                      &        &       &      &  $<10$  &  2719.901  &   $-1.9$  &  9.7 & RV = $-2.0$       \\
\hline
\end{tabular}
\begin{list}{}
\item Table~\ref{spectro} Cont/d.
\end{list}
\end{table}

\newpage


   \begin{table}
      \caption[]{A spectral-line data ensemble, for each star detailed 
              in Table~\ref{inputcat}, is presented for the Li {\sc i} 
              6708\r{A} \& Balmer H$\alpha$ 6563\r{A} lines .}
         \label{Li-table}
\begin{tabular}{llrllllll}
\hline\hline
TARGET              &  HJD       & S/N$^{a}$ & Li 6707.8    & Li 6707.8   &  Li 6707.8   & Li 6707.8   & H$\alpha$  & H$\alpha$  \\
                    & (+2450000) &           & EW [\r{A}]    & EW [\r{A}]   &  EW [\r{A}]   & EW [\r{A}]  & EW [\r{A}]  & EW [\r{A}]  \\ 
                    &  [days]    &           & (INT$^{b}$)  &  (GF$^{c}$) &  (ADP$^{d}$) & (LIT$^{e}$) & (GF$^{c}$) & (INT$^{b}$) \\ \hline
Lupus :             &            &           &          &          &                     &        &         &       \\ \hline
RX J1507.2-3505     &  2713.762  &  50       &   0.319  &  0.330   &  $0.347 \pm 0.035$  & 0.324  &  0.622  & 0.634 \\
                    &  2718.766  &  74       &   0.350  &  0.351   &                     &        &  0.553  & 0.586 \\
RX J1507.4-4601     &  2718.790  &  62       &   0.326  &  0.336   &  $0.336 \pm 0.034$  &        &  0.637  & 0.831 \\
RX J1515.7-3331     &  2713.860  &  77       &   0.376  &  0.410   &  $0.385 \pm 0.038$  & 0.358  &  0.871  & 0.904 \\
                    &  2718.925  &  80       &   0.353  &  0.364   &                     &        &  0.745  & 0.923 \\
RX J1518.4-3738     &  2718.746  &  90       &   0.352  &  0.365   &  $0.365 \pm 0.037$  & 0.347  &  0.673  & 0.970 \\
RX J1523.4-4055     &  2719.769  &  22       &   0.399  &  0.408   &  $0.408 \pm 0.041$  & 0.391  &  1.057  & 1.115 \\
RX J1524.5-3652     &  2713.826  &  34       &   0.375  &  0.402   &  $0.389 \pm 0.039$  & 0.351  &  0.809  & 0.690 \\
                    &  2719.792  &  36       &   0.377  &  0.379   &                     &        &         &       \\
RX J1525.0-3604     &  2718.852  &  82       &   0.372  &  0.385   &  $0.385 \pm 0.039$  &        &  0.648  & 0.797 \\
RX J1529.6-3546     &  2718.879  &  92       &   0.401  &  0.413   &  $0.413 \pm 0.041$  &        &  0.672  & 1.072 \\
                    &  2718.879  &           &          &          &                     &        &  0.754  & 1.198 \\
RX J1529.7-3628     &  2719.715  &  25       &   0.443  &  0.438   &  $0.438 \pm 0.044$  &        &  1.518  & 1.878 \\
RX J1546.6-3618     &  2713.705  &  39       &   0.372  &  0.376   &  $0.376 \pm 0.038$  & 0.381  &  1.605  & 1.914 \\
                    &  2719.747  &           &          &          &                     &        &         &       \\
RX J1547.6-4018     &  2713.737  &  38       &  0.390   &  0.395   &  $0.376 \pm 0.037$  & 0.385  &  0.717  & 0.903 \\
                    &  2718.916  &  67       &  0.368   &  0.374   &                     &        &  0.768  & 1.138 \\
RX J1601.1-3320     &  2713.799  &  68       &  0.348   &  0.352   &  $0.352 \pm 0.035$  & 0.308  &  0.822  & 0.786 \\
RX J1605.7-3905     &  2713.780  &  43       &          &  0.350   &  $0.350 \pm 0.035$  &        &  0.721  & 0.976 \\
RX J1608.9-3905     &  2718.900  & 100       &  0.455   &  0.483   &  $0.483 \pm 0.049$  & 0.491  &  0.593  & 1.147 \\
                    &  2718.900  &           &          &          &                     &        &  0.721  & 1.186 \\ \hline
Chamaeleon {\sc i} $\&$ {\sc ii}   &  &        &          &          &                     &        &          &       \\ \hline
RX J0850.1-7554     &  2712.590  &  87       &  0.248   &  0.278   &  $0.283 \pm 0.028$  & 0.25   &  0.497  & 0.493 \\
                    &  2718.521  &  62       &  0.265   &  0.300   &                     &        &         &       \\
                    &  2718.521  &           &          &          &                     &        &         &       \\

RX J0951.9-7901     &  2712.620  &  98       &  0.353   &  0.405   &  $0.405 \pm 0.041$  & 0.26   &  0.857  & 1.295 \\
                    &  2712.620  &           &          &          &                     &        &         &       \\
                    &  2718.533  &  72       &  0.397   &  0.457   &  $0.457 \pm 0.046$  &        &  0.661  & 0.902 \\
                    &  2718.533  &           &          &          &                     &        &  0.650  & 1.054 \\
RX J1112.7-7637     &  2712.694  &  64       &  0.452   &  0.482   &  $0.470 \pm 0.047$  & 0.47   &  1.707  & 2.178 \\
                    &  2718.546  &  49       &  0.414   &  0.435   &                     &        &  0.693  & 0.909 \\
RX J1129.2-7546     &  2713.628  &  27       &  0.484   &  0.497   &  $0.499 \pm 0.050$  & 0.45   &  0.804  & 0.852 \\
                    &  2719.528  &  11       &  0.527   &  0.570   &                     &        &         &       \\
RX J1140.3-8321     &  2718.563  &  64       &  0.195   &  0.198   &  $0.198 \pm 0.020$  & 0.21   &  0.893  & 1.000 \\
RX J1158.5-7754a    &  2713.656  &  65       &  0.495   &  0.501   &  $0.483 \pm 0.048$  & 0.48   &  1.384  & 1.268 \\ 
                    &  2718.614  &  87       &  0.473   &  0.478   &                     &        &  1.150  & 1.326 \\
RX J1159.7-7601     &  2718.636  &  90       &  0.430   &  0.445   &  $0.445 \pm 0.045$  & 0.50   &  0.892  & 1.147 \\
RX J1201.7-7859     &  2712.734  & 100       &  0.252   &  0.259   &  $0.260 \pm 0.026$  & 0.24   &  0.313  & 0.307 \\
                    &  2718.678  & 160       &  0.246   &  0.260   &                     &        &  0.283  & 0.268 \\
RX J1233.5-7523     &  2712.760  &  81       &  0.151   &  0.144   &  $0.135 \pm 0.013$  & 0.13   &  0.142  & 0.136 \\
                    &  2718.716  &  95       &  0.127   &  0.130   &                     &        &  0.132  & 0.122 \\
RX J1239.4-7502     &  2713.675  &  81       &  0.426   &  0.436   &  $0.459 \pm 0.046$  & 0.40   &  0.660  & 0.759 \\
                    &  2718.726  &  82       &  0.452   &  0.480   &                     &        &  0.771  & 0.915 \\
RX J1303.5-7701     &  2712.779  &           &          &          &                     &        &         &       \\
RX J1307.3-7708     &  2712.794  &   8       &  0.082   &  0.095   &  $0.095 \pm 0.010$  &        &         &       \\ \hline
\end{tabular}
\vspace*{1mm}
\begin{list}{}
\item ${\mathrm{a}}~$ Approximate S/N of nearest continuum region to the blaze centre 
of the \lii 6708 \r{A} order. 
\item ${\mathrm{b}}~$ EWs measured using direct integration between the limits of 
where the \lii/H$\alpha$ feature approaches unity on either side of its central 
absorption/residual emission feature.  
\item ${\mathrm{c}}~$ EWs measured using a Gaussian fit, with the same wavelength 
limits as for the direct integration method.
\item ${\mathrm{d}}~$ Adopted \lii 6708\r{A} EW, using weighted mean of the Gaussian 
fitted EWs for targets having spectra recorded on different observing nights. 
Errors levels are estimated to be $10\%$.
\item ${\mathrm{e}}~$ \lii 6708\r{A} EWs for these targets as reported in the extant 
literature. 
\end{list}
%
\end{table}

\newpage

   \begin{table}
\begin{tabular}{llrllllll}
\hline\hline
TARGET              &  HJD       & S/N$^{a}$ & Li 6707.8    & Li 6707.8   &  Li 6707.8   & Li 6707.8   & H$\alpha$  & H$\alpha$  \\
                    & (+2450000) &           & EW [\r{A}]    & EW [\r{A}]   &  EW [\r{A}]   & EW [\r{A}]  & EW [\r{A}]  & EW [\r{A}]  \\ 
                    &  [days]    &           & (INT$^{b}$)  &  (GF$^{c}$) &  (ADP$^{d}$) & (LIT$^{e}$) & (GF$^{c}$) & (INT$^{b}$) \\ \hline
Rho Ophiuchus :     &            &           &          &          &                     &        &         &       \\ \hline
RX J1627.1-2419     &  2712.852  &  14       &  0.222   &  0.275   &  $0.275 \pm 0.028$  & 0.16   &  0.559  & 1.208 \\
RX J1625.6-2613     &  2712.886  &  42       &  0.512   &  0.500   &  $0.500 \pm 0.050$  & 0.45   &  4.287  & 5.075 \\ \hline
CrA SFR :           &            &           &                  &          &                     &         &         & \\ \hline
CrAPMS  2           &  2713.909  &  98       &  0.334           &  0.353   &  $0.352 \pm 0.035$  & 0.28    &  0.987  & 1.031 \\
                    &  2719.923  &  46       &  0.315           &  0.329   &                     &         &  0.711  & 0.982 \\
RX J1839.0-3726     &  2719.869  &  45       &  0.385           &  0.395   &  $0.395 \pm 0.040$  & 0.34    &  0.692  & 0.870 \\
RX J1853.1-3609     &  2719.837  &  58       &  0.378           &  0.394   &  $0.394 \pm 0.040$  & 0.39    &  0.592  & 0.730 \\
RX J1917.4-3756     &  2713.920  &  93       &  0.471           &  0.490   &  $0.489 \pm 0.049$  & 0.48    &  1.049  & 1.110 \\
                    &  2719.817  &  48       &  0.457           &  0.472   &                     &         &  0.747  & 1.001 \\
CrAPMS  4SE         &  2719.901  &  51       &  0.372           &  0.377   &  $0.377 \pm 0.038$  & 0.36    &  0.584  & 1.041 \\
\hline
\end{tabular}
\begin{list}{}
\item Table~\ref{Li-table} Cont/d.
\end{list}
\end{table}

\newpage 

   \begin{table}
      \caption[]{Fundamental stellar parameters, and the intermediate data used to 
             calculate them, are presented for each SFR candidate member. 
             Mass and Age estimates are derived by comparison of their 
             luminosity-temperature data with the theoretical models and 
             tracks computed by D'Antona and Mazzitelli (1997).}
         \label{Ltab}
%
%
%
%
%
%
%
%
%
%
%
%
%
\begin{tabular}{lccrccccccrcc}
\hline\hline
TARGET  & V & V$-$I& Sp.  & V$-$I$^{a}$ & J$-$H$^{a}$ & E$_{V-I}^{b}$ &  Av$^{c}$ &  B.C.$^{d}$ & T$_{eff}^{d}$ & log(L$_{bol}$/ & Mass & Age  \\ 
        &   &      & Type & KH95  & KH95  & or E$_{J-H}$     & [mags]& [mags] &  [K]      &     ~~~~ L$_{\odot}$)$^{e}$  & [M$_{\odot}$] & [Myr] \\ \hline
Lupus :          &        &      &              &       &           &       &       &       &         &        &     \\ \hline
RX J1507.2-3505  &  10.53 & 1.26 & K0   & 0.85  &       & 0.41      & 1.02  & -0.31 & 5250  &  0.500  & 1.8    & 3.0 \\
RX J1507.4-4601  &  11.71 &      & K1   & 0.93  & 0.47  & 0.00      & 0.00  & -0.37 & 5080  & -0.357  & 0.9    & 28  \\
RX J1515.7-3331  &  10.69 & 1.00 & K0   & 0.85  &       & 0.15      & 0.37  & -0.31 & 5250  &  0.175  & 1.4    & 8.0 \\
RX J1518.4-3738  &  10.92 & 1.01 & K1   & 0.93  &       & 0.08      & 0.20  & -0.37 & 5080  &  0.039  & 1.2    & 6.5 \\
RX J1523.4-4055  &  11.87 & 1.16 & K2   & 1.01  &       & 0.15      & 0.37  & -0.42 & 4900  & -0.253  & 1.0    & 14  \\ \hline
RX J1524.5-3652  &  11.30 & 1.01 & K1   & 0.93  &       & 0.08      & 0.20  & -0.37 & 5080  & -0.113  & 1.1    & 12  \\
RX J1525.0-3604  &  10.92 & 1.09 & K1   & 0.93  &       & 0.16      & 0.40  & -0.37 & 5080  &  0.119  & 1.3    & 6.0 \\
RX J1529.6-3546  &        &      & K1   & 0.93  &       &           &       & -0.37 & 5080  &         &        &     \\
RX J1529.7-3628  &        &      & K2   & 1.01  &       &           &       & -0.42 & 4900  &         &        &     \\
RX J1546.6-3618  &  11.28 & 1.07 & K1   & 0.93  &       & 0.14      & 0.35  & -0.37 & 5080  & -0.045  & 1.2    & 8.5 \\
RX J1547.6-4018  &  11.08 & 1.05 & K1   & 0.93  &       & 0.12      & 0.30  & -0.37 & 5080  &  0.015  & 1.2    & 7.0 \\
RX J1601.1-3320  &  10.88 & 1.07 & G8   & 0.81  &       & 0.26      & 0.64  & -0.25 & 5520  &  1.525  & $>2.5$ & $<1.0$  \\ 
RX J1605.7-3905  &  10.49 & 0.90 & G9   & 0.83  &       & 0.07      & 0.17  & -0.28 & 5410  &  0.163  & 1.3    & 11    \\
RX J1608.9-3905  &  10.88 & 1.14 & K2   & 1.01  &       & 0.13      & 0.32  & -0.42 & 4900  &  0.123  & 1.3    & 4.0   \\ \hline
Chamaeleon {\sc i} $\&$ {\sc ii}   &  &  &  &     &       &           &       &       &       &         &        &        \\ \hline
RX J0850.1-7554  &  10.57 & 0.82 & G6   & 0.77  &       & 0.05      & 0.12  & -0.22 & 5700  &  0.203  & 1.2    & 17    \\
RX J0951.9-7901  &  10.22 & 0.93 & G7   & 0.79  &       & 0.14      & 0.35  & -0.23 & 5630  &  0.439  & 1.5    & 8.0   \\
RX J1112.7-7637  &  12.43 & 1.41 & K0/4 & 1.01  &       & 0.40      & 0.99  & -0.42 & 4900  & -0.113  & 1.1    & 7.5   \\
RX J1129.2-7546  &  12.91 & 1.76 & K3   & 1.08  &       & 0.68      & 1.69  & -0.50 & 4730  &  0.007  & 1.0    & 3.0   \\
RX J1140.3-8321  &  11.56 & 1.33 & K3/4 & 1.08  &       & 0.25      & 0.62  & -0.50 & 4730  &  0.119  & 1.0    & 2.0   \\
RX J1158.5-7754a &  10.59 & 1.43 & K2/4 & 1.08  &       & 0.35      & 0.87  & -0.50 & 4730  &  0.607  & 1.0    & 0.5   \\
RX J1159.7-7601  &  11.30 & 1.37 & K2/4 & 1.08  &       & 0.29      & 0.72  & -0.50 & 4730  &  0.263  & 1.0    & 1.0   \\
RX J1201.7-7859  &   8.59 &      & G5   & 0.76  & 0.36  & 0.00      & 0.00  & -0.21 & 5770  &  0.943  & 2.1    & 3.0   \\
RX J1233.5-7523  &   9.47 &      & K1   & 0.93  & 0.47  & 0.00      & 0.00  & -0.37 & 5080  &  0.655  & 1.8    & 0.5   \\
RX J1239.4-7502  &  10.24 & 1.06 & K2/3 & 1.01  &       & 0.05      & 0.12  & -0.42 & 4900  &  0.415  & 1.4    & 2.0   \\ 
RX J1303.5-7701  &        &      & G7   & 0.79  &       &           &       & -0.23 & 5630  &         &        &       \\ 
RX J1307.3-7708  &        &      & G6   & 0.77  &       &           &       & -0.22 & 5700  &         &        &       \\ \hline
Rho Ophiuchus :  &        &      &      &       &       &           &       &       &       &         &        &       \\ \hline
RX J1627.1-2419  &        &      & G1   & 0.72  &       &           &       & -0.19 & 5945  &         &        &       \\
RX J1625.6-2613  &        &      & G8   & 0.81  &       &           &       & -0.25 & 5520  &         &        &       \\ \hline
CrA SFR :        &        &      &      &       &       &           &       &       &       &         &        &       \\ \hline
CrAPMS  2        &  10.44 & 0.93 & G5   & 0.76  &       & 0.17      & 0.42  & -0.21 & 5770  &  1.323  & $>2.5$ &  $<5.0$ \\ 
RX J1839.0-3726  &  10.81 & 1.00 & K1   & 0.93  &       & 0.07      & 0.17  & -0.37 & 5080  & -0.063  & 1.1    &  9.0   \\
RX J1853.1-3609  &   9.60 &      & K2   & 1.01  & 0.50  & 0.00      & 0.00  & -0.42 & 4900  &         &        &        \\
RX J1917.4-3756  &   9.90 &      & K2   & 1.01  & 0.50  & 0.04      & 0.35  & -0.42 & 4900  &  0.393  & 1.4    &  2.0   \\
CrAPMS  4SE      &  10.95 & 1.00 & G5   & 0.76  &       & 0.24      & 0.60  & -0.21 & 5770  & -0.011 & $1.0-1.2$ & 20.0 \\
\hline
\end{tabular}
\begin{list}{}
\item ${\mathrm{a}}~$ Theoretical V$-$I or J$-$H values for each target's spectral type 
(assumes dwarf-class) are taken from KH95. 
\item ${\mathrm{b}}~$ E$_{V-Ic}$ and E$_{J-H}$ are calculated by subtracting the 
target's V$-$Ic or J$-$H colour from their theoretical values based on spectral type. 
\item ${\mathrm{c}}~$ We assume Av = $2.48\times$E$_{(V-I)c}$ or 
Av = $8.38\times$E$_{J-H}$. 
\item ${\mathrm{d}}~$ Bolometric corrections [B.C.] and effective temperatures, as 
a function of spectral type, are taken from KH95.
\item ${\mathrm{e}}~$ The luminosity of the Sun is assumed to be $3.85\times 10^{33}$ 
erg~s$^{-1}$.
\end{list}
\end{table}

%
%
%

\begin{table}
\caption[]{Addition Fe {\sc ii} lines used in the metallicity analysis presented herein, 
to complement those used in an identical analysis procedure performed by 
Santos et al. (2004). }
\begin{tabular}{lccc}
\hline\hline
   $\lambda$ (\r{A}) & $\chi_l$ & $\log{gf}$ & EW$_{\sun}$ [m\r{A}] \\
\hline
   5325.56  & 3.220  & -3.20 & 41.4\\
   5414.07  & 3.220  & -3.57 & 26.9\\
   5425.25  & 3.200  & -3.23 & 41.5\\
   6456.39  & 3.900  & -2.11 & 62.9\\
\hline
\end{tabular}
\label{tab:feii}
\end{table}

\begin{table}
\caption{Stellar parameters, determined from a spectroscopic metallicity 
analysis of Fe lines in spectra of suitable WTTS candidate members, are 
presented for each of the SFRs under study. The effective temperatures 
listed in column 8 are derived from a spectral-type versus temperature 
analysis (see Figure~4 and de Jager \& 
Nieuwenhuijzen 1987).}
\begin{tabular}{lccccccc}
\hline\hline
\multicolumn{1}{c}{Star} & $ T_{eff} $ & $\log g $          & $ V_t $          & $ [Fe/H] $ & $ N $ 
(Fe {\sc i}, Fe {\sc ii}) & $ \sigma $ (Fe {\sc i}, Fe {\sc ii}) & T$_{eff}$ \\
                     &  [K]      & [cm s$ ^{-2}$ ]    & [km s$ ^{-1} $ ] &            &       
                       &                            & [K]        \\
\hline
\multicolumn{7}{c}{Chamaeleon}\\
\hline
RXJ1140.3-8321  & $ 4814 \pm 154$  & $4.07 \pm 0.92 $  &  $2.15 \pm 0.23 $   &  $-0.14 \pm  0.15 $  & $23,4$  & $0.14,0.51$  & $4688 \pm 177$  \\ 
RXJ1158.5-7754a & $ 4810 \pm 142$  & $3.87 \pm 0.51 $  &  $2.11 \pm 0.22 $   &  $-0.26 \pm  0.15 $  & $20,2$  & $0.13,0.25$  & $4688 \pm 177$  \\ 
RXJ1159.7-7601  & $ 4836 \pm 146$  & $4.23 \pm 0.33 $  &  $2.07 \pm 0.22 $   &  $-0.18 \pm  0.16 $  & $29,6$  & $0.16,0.12$  & $4688 \pm 177$  \\ 
RXJ1201.7-7859  & $ 5902 \pm 163$  & $4.38 \pm 0.43 $  &  $1.99 \pm 0.31 $   &  $0.12 \pm  0.17 $  & $14,4$  & $0.09,0.18$   & $5555 \pm 178$  \\ 
RXJ1233.5-7523  & $ 5469 \pm ~62$  & $4.40 \pm 0.22 $  &  $1.52 \pm 0.10 $   &  $-0.09 \pm  0.08 $  & $35,10$  & $0.07,0.10$ & $4989 \pm 183$  \\ 
\hline
\multicolumn{7}{c}{Lupus}\\
\hline
RXJ1507.2-3505  & $ 5155 \pm 148$  & $4.37 \pm 0.49 $  &  $2.53 \pm 0.26 $   &  $-0.08 \pm  0.14 $  & $21,7$  & $0.11,0.23$  & $5152 \pm 188$  \\ 
RXJ1546.6-3618  & $ 5062 \pm 109$  & $4.09 \pm 0.53 $  &  $2.12 \pm 0.16 $   &  $-0.12 \pm  0.14 $  & $34,6$  & $0.13,0.26$  & $4989 \pm 183$  \\ 
RXJ1547.6-4018  & $ 5045 \pm 120$  & $4.18 \pm 0.61 $  &  $2.23 \pm 0.20 $   &  $-0.14 \pm  0.13 $  & $30,8$  & $0.11,0.31$  & $4989 \pm 183$  \\ 
RXJ1601.1-3320  & $ 5479 \pm 152$  & $3.70 \pm 0.35 $  &  $2.49 \pm 0.26 $   &  $-0.05 \pm  0.18 $  & $24,3$  & $0.14,0.15$  & $5309 \pm 183$  \\ 
\hline
\multicolumn{7}{c}{CrA}\\
\hline
CrAPMS2         & $ 5400 \pm 164$  & $4.06 \pm 0.45 $  &  $2.41 \pm 0.30 $   &  $-0.09 \pm  0.18 $  & $21,5$  & $0.13,0.19$  & $5555 \pm 178$  \\ 
CrAPMS4SE       & $ 5335 \pm 101$  & $4.59 \pm 0.41 $  &  $2.24 \pm 0.17 $   &  $-0.02 \pm  0.12 $  & $31,10$  & $0.10,0.20$ & $5555 \pm 178$  \\ 
RXJ1839.0-3726  & $ 5293 \pm 127$  & $4.51 \pm 0.42 $  &  $2.19 \pm 0.21 $   &  $-0.01 \pm  0.15 $  & $35,9$  & $0.14,0.19$  & $4989 \pm 183$  \\ 
\hline
\end{tabular}
\label{tab:parameters}
\end{table}

\begin{table}
\caption[]{Average and {\sc rms} of the metallicities derived for each star 
formation region. $N$ is the number of stars with [Fe/H] values in each case.}
\begin{tabular}{lccc}
\hline\hline
   Region & $N$ & $<$[Fe/H]$>$  & {\sc rms} \\
\hline
\multicolumn{4}{l}{All stars}\\
Cha  & 5 & $-$0.11 & 0.14\\
Lup  & 4 & $-$0.10 & 0.04\\
CrA  & 3 & $-$0.04 & 0.05\\ \hline
\multicolumn{4}{l}{Stars with $v\sin{i}<15$km~s$^{-1}$}\\ \hline
Cha  & 4 & $-$0.17 & 0.07\\
Lup  & 3 & $-$0.11 & 0.03\\
CrA  & 2 & $-$0.02 & 0.00\\
\hline
\end{tabular}
\label{tab:avgfeh}
\end{table}

\end{document}